\documentclass{mn2e}
\usepackage{graphicx}

\def \xoff {\ifmmode x_{\rm off} \else $x_{\rm off}$ \fi}
\def \rhorms {\ifmmode \rho_{\rm rms} \else $\rho_{\rm rms}$ \fi}
\def \rhocrit {\ifmmode \rho_{\rm crit} \else $\rho_{\rm crit}$ \fi}

\def \mnras {MNRAS}
\def \etal {et~al.~}
\def \chisq  {\ifmmode  \chi^2   \else  $\chi^2$  \fi}  
\def \chisqr {\ifmmode \chi^2_{\rm r} \else $\chi^2_{\rm r}$ \fi}
\def \spose#1{\hbox  to 0pt{#1\hss}}  
\def \lta{\mathrel{\spose{\lower 3pt\hbox{$\sim$}}\raise  2.0pt\hbox{$<$}}}
\def \gta{\mathrel{\spose{\lower  3pt\hbox{$\sim$}}\raise 2.0pt\hbox{$>$}}}
 
\def \ha  {\ifmmode H\alpha \else H$\alpha $ \fi}

\def \kms {\ifmmode  \,\rm km\,s^{-1} \else $\,\rm km\,s^{-1}  $ \fi }
\def \kpc {\ifmmode  {\rm kpc}  \else ${\rm  kpc}$ \fi  }  
\def \Msun {\ifmmode M_{\odot} \else $M_{\odot}$ \fi} 
\def \hMsun {\ifmmode h^{-1}\,\rm M_{\odot} \else $h^{-1}\,\rm M_{\odot}$ \fi}
\def \hhMsun {\ifmmode h^{-2}\,\rm M_{\odot}\else $h^{-2}\,\rm M_{\odot}$ \fi}
\def \Lsun {\ifmmode L_{\odot} \else $L_{\odot}$ \fi} 
\def \hhLsun {\ifmmode h^{-2}\,\rm L_{\odot} \else $h^{-2}\,\rm L_{\odot}$ \fi}

\def \LCDM {\ifmmode \Lambda{\rm CDM} \else $\Lambda{\rm CDM}$ \fi}
\def \sig8 {\ifmmode \sigma_8 \else $\sigma_8$ \fi} 
\def \OmegaM {\ifmmode \Omega_{\rm M} \else $\Omega_{\rm M}$ \fi} 
\def \OmegaL {\ifmmode \Omega_{\rm \Lambda} \else $\Omega_{\rm \Lambda}$\fi} 
\def \Deltavir {\ifmmode \Delta_{\rm vir} \else $\Delta_{\rm vir}$ \fi}

\def \rs {\ifmmode r_{\rm s} \else $r_{\rm s}$ \fi} 
\def \rrm2 {\ifmmode r_{-2} \else $r_{-2}$ \fi} 
\def \ccm2 {\ifmmode c_{-2} \else$c_{-2}$ \fi} 
\def \cvir {\ifmmode c_{\rm vir} \else $c_{\rm vir}$ \fi} 
\def \cbar {\ifmmode \overline{c} \else $\overline{c}$ \fi}

\def \R200 {\ifmmode R_{200} \else $R_{200}$ \fi} 
\def \Rvir {\ifmmode R_{\rm vir} \else $R_{\rm vir}$ \fi}

\def \v200 {\ifmmode V_{200} \else $V_{200}$ \fi} 
\def \Vvir {\ifmmode V_{\rm  vir} \else  $V_{\rm vir}$  \fi} 
\def  \Vhalo  {\ifmmode V_{\rm halo} \else $V_{\rm halo}$ \fi}

\def \M200 {\ifmmode M_{200} \else $M_{200}$ \fi} 
\def \Mvir {\ifmmode M_{\rm  vir} \else $M_{\rm  vir}$ \fi}  
\def \Mshell  {\ifmmode M_{\rm shell} \else $M_{\rm shell}$ \fi}

\def \Nvir {\ifmmode N_{\rm  vir} \else $N_{\rm  vir}$ \fi}  

\def \Jvir {\ifmmode J_{\rm vir} \else $J_{\rm vir}$ \fi} 
\def \Jshell {\ifmmode J_{\rm shell} \else $J_{\rm shell}$ \fi}

\def \Evir {\ifmmode E_{\rm vir} \else $E_{\rm vir}$ \fi} 

\def \lam {\ifmmode \lambda  \else $\lambda$ \fi} 
\def \lamp {\ifmmode \lambda^{\prime} \else $\lambda^{\prime}$  \fi} 
\def \lampc {\ifmmode \lambda^{\prime}_{\rm c} \else
  $\lambda^{\prime}_{\rm c}$  \fi} 
\def \lambar {\ifmmode \bar{\lambda}  \else  $\bar{\lambda}$  \fi}  
\def  \lampbar  {\ifmmode \bar{\lambda^{\prime}} \else
  $\bar{\lambda^{\prime}}$\fi} 
\def \siglam {\ifmmode \sigma_{\lambda} \else $\sigma_{\lambda}$ \fi} 
\def \siglamp {\ifmmode                \sigma_{\lambda^{\prime}} \else
$\sigma_{\lambda^{\prime}}$\fi}

\def \Rd {\ifmmode R_{\rm d} \else $R_{\rm d}$ \fi} 
\def \Rs {\ifmmode R_{\rm s} \else $R_{\rm s}$ \fi}  
\def \Rd {\ifmmode R_{\rm d} \else $R_{\rm d}$ \fi}  
\def \Rcool  {\ifmmode R_{\rm  cool}  \else $R_{\rm cool}$ \fi} 
\def \RIII {\ifmmode  3.2\Rs \else $3.2\Rs$ \fi} 
\def \RII {\ifmmode 2.2\Rs \else $2.2\Rs$  \fi} 
\def \Reff {\ifmmode R_{\rm eff} \else $R_{\rm  eff}$ \fi} 
\def  \rb {\ifmmode r_{\rm b}  \else $r_{\rm b}$ \fi}

\def  \Sigmacrit   {\ifmmode  \Sigma_{\rm  crit}   
\else  $\Sigma_{\rm crit}$\fi} 
\def \Sig0 {\ifmmode \Sigma_{0} \else $\Sigma_{0}$ \fi}

\def \muI {\ifmmode \mu_{0,I} \else $\mu_{0,I}$ \fi}

\def \mgal {\ifmmode m_{\rm gal} \else $m_{\rm gal}$ \fi} 
\def \md {\ifmmode m_{\rm d} \else $m_{\rm d}$ \fi} 
\def \ms {\ifmmode m_{\rm   s}   \else   $m_{\rm   s}$   \fi}   
\def   \mdbar   {\ifmmode {\overline{m}}_{\rm d} \else
  ${\overline{m}}_{\rm d}$ \fi} 
\def \msbar {\ifmmode  \bar{m}_{\rm  s}  \else  $\bar{m}_{\rm s}$
  \fi}  
\def  \Md {\ifmmode M_{\rm d}  \else $M_{\rm d}$ \fi} 
\def  \Ms {\ifmmode M_{\rm s} \else $M_{\rm  s}$ \fi} 
\def \Mb {\ifmmode  M_{\rm b} \else $M_{\rm b}$ \fi} 
\def \Mstar {\ifmmode  M_{\rm star} \else $M_{\rm star}$ \fi}
\def \Mdisc {\ifmmode M_{\rm disc} \else $M_{\rm disc}$ \fi}

\def \Jd {\ifmmode J_{\rm d} \else $J_{\rm d}$ \fi} 
\def \Jb {\ifmmode J_{\rm b} \else $J_{\rm b}$ \fi}  
\def \fb {\ifmmode  f_{\rm b} \else $f_{\rm b}$ \fi}

\def  \jd  {\ifmmode j_{\rm  d}  \else  $j_{\rm  d}$ \fi}  
\def  \jdmd {\ifmmode \frac{j_{\rm  d}}{m_{\rm d}} \else
  $\frac{j_{\rm d}}{m_{\rm d}}$ \fi} 
\def \fj {\ifmmode f_{\rm j} \else $f_{\rm j}$ \fi} 
\def \ft {\ifmmode f_{\rm t}  \else $f_{\rm t}$ \fi} 
\def  \fM {\ifmmode f_{\rm M} \else $f_{\rm M}$ \fi}

\def  \Vd {\ifmmode  V_{\rm  d}  \else $V_{\rm  d}$  \fi} 
\def  \Vcool {\ifmmode V_{\rm cool} \else $V_{\rm cool}$ \fi} 
\def \Vcirc {\ifmmode V_{\rm circ}  \else $V_{\rm circ}$  \fi} 
\def \VIII  {\ifmmode V_{3.2} \else $V_{3.2}$ \fi} 
\def  \VII {\ifmmode V_{2.2} \else $V_{2.2}$ \fi}
\def \Vobs {\ifmmode V_{\rm obs}  \else $V_{\rm obs}$ \fi} 
\def \Vdisc {\ifmmode V_{\rm disc} \else  $V_{\rm disc}$ \fi} 
\def \Vmax {\ifmmode V_{\rm  max} \else  $V_{\rm max}$  \fi} 
\def  \Vmaxobs{\ifmmode V_{\rm max}^{\rm obs}\else  $V_{\rm max}^{\rm
    obs}$\fi}  
\def \Vtot {\ifmmode V_{\rm tot} \else $V_{\rm tot}$  \fi} 
\def \Vrot {\ifmmode V_{\rm rot} \else  $V_{\rm rot}$  \fi} 
\def  \Vflat {\ifmmode  V_{\rm  flat} \else $V_{\rm flat}$ \fi}

\def \Ups {\ifmmode \Upsilon  \else $\Upsilon$ \fi} 
\def \YB {\ifmmode \Upsilon_B \else $\Upsilon_B$ \fi} 
\def \YI {\ifmmode  \Upsilon_I  \else $\Upsilon_I$ \fi} 
\def \DeltaIMF {\ifmmode \Delta_{\rm IMF} \else $\Delta_{\rm IMF}$ \fi}

\def\LCDM{$\Lambda$CDM }

\def\c200{$c_{200}$}

\newcommand{\mpch}{\>h^{-1}{\rm {Mpc}}}

\title[Properties of DM Haloes] {Properties of Dark Matter Haloes
and their Correlations: the Lesson from Principal Component Analysis}
\author[R. A. Skibba \& A. V.  Macci\`o ]  {Ramin A. Skibba$^{1,2}$\thanks{rskibba@as.arizona.edu}, Andrea V. Macci\`o$^{1}$\thanks{maccio@mpia.de} \\ 
$^1$Max-Planck-Institut f\"ur Astronomie, K\"onigstuhl 17, 69117 Heidelberg, Germany \\
$^2$Steward Observatory, University of Arizona, 933 N. Cherry Ave., Tucson, AZ 85721, USA
}

\newcounter{appfig}

\begin{document}
             
\date{submitted to MNRAS}
             

\maketitle           

\label{firstpage}
             
\begin{abstract}

We study the correlations between the structural parameters of dark matter haloes 
using Principal Component Analysis (PCA). 
We consider a set of eight parameters, six of which are commonly used to characterize 
dark matter halo properties:  mass, concentration, spin, shape, overdensity, 
and the angle ($\Phi_L$) between the major axis and the angular momentum vector. 
Two additional parameters ($\xoff$ and $\rhorms$) are used to describe 
the degree of `relaxedness' of the halo. 
We find that we can account for much of the variance of these properties 
with halo mass and concentration, on the one hand, and halo relaxedness 
on the other.  Nonetheless, three principle components are usually 
required to account for most of the variance.  
We argue that halo mass is not as dominant as expected, 
which is a challenge for halo occupation models and semi-analytic models 
that assume that mass determines other halo (and galaxy) properties. 
In addition, we find that the angle $\Phi_L$ is not significantly 
correlated with other halo parameters, which may present a difficulty 
for models in which galaxy disks are oriented in haloes in a particular way.
Finally, at fixed mass, we find that a halo's environment (quantified by the 
large-scale overdensity) is relatively unimportant.

\end{abstract}


\setcounter{footnote}{1}

\section{Introduction}
\label{sec:intro}

In the paradigm of hierarchical structure formation, gravitational evolution 
causes dark matter particles to cluster around peaks of the initial density 
field and to collapse into virialized objects (haloes), which provide the 
potential wells in which galaxies subsequently form (White \& Rees 1978).  It is therefore expected 
that the properties of a galaxy are correlated with the properties of its host 
halo.  Small haloes merge to form larger and more massive haloes, which tend to be 
located in dense environments and are expected to host groups of galaxies.

It is commonly assumed that the mass of a dark matter halo determines how the 
galaxies it hosts form and evolve; other halo properties are assumed to be 
less important at fixed mass.  This assumption is usually made both by those who 
use halo models of galaxy clustering (e.g., Zehavi et al.\ 2005; Skibba et al.\ 
2006; van den Bosch et al.\ 2007; Moster et al.\ 2010) and by those who use semi-analytic models of 
galaxy formation (e.g., Bower et al.\ 2006; De Lucia \& Blaizot 2007; Cattaneo 
et al.\ 2007; Somerville et al.\ 2008).  Many of such models have recently been able to provide a 
description, and perhaps a plausible explanation, of various galaxy statistics, 
such as the luminosity function, stellar mass function, colour-magnitude 
distribution, correlation function, and mark correlation functions, lending 
credence to claims about the importance of halo mass.

Nonetheless, although halo properties, such as their mass, spin, concentration, 
and shape are correlated with each other, there is considerable scatter in these 
correlations (e.g., Avila-Reese et al.\ 2005; Macci\`{o} et al.\ 2007; Macci\`{o} 
et al.\ 2008; Ragone-Figueroa et al.\ 2010). 
In addition, recent studies of numerical simulations have shown 
that halo formation time is correlated with the environment at fixed mass (e.g., 
Sheth \& Tormen 2004; Gao et al.\ 2005; Dalal et al.\ 2008).  Some have argued that the halo 
occupation distribution (used in models of galaxy clustering) is correlated with 
formation time at fixed mass (Zentner et al.\ 2005a; Wechsler et al.\ 2006; Giocoli et al.\ 2010). 
It has also been argued that such halo `assembly bias' affects the clustering of 
galaxies and that the effect varies with luminosity and colour (Croton et al.\ 
2007).  Nonetheless, careful studies of the environmental dependence of galaxy 
luminosity and colour in the real universe have explained this dependence in 
terms of the environmental dependence of halo mass alone (e.g., Tinker et al.\ 
2008a; Skibba \& Sheth 2009). 
Such models are also encouraged by an analysis of a 
halo-based galaxy group catalogue, in which, given a halo's mass, the colours of 
the galaxies it hosts are not significantly correlated with the large-scale 
density field (Blanton \& Berlind 2007). 
In addition, the halo model can reproduce the observed non-monotonic relation 
between clustering strength and density (Abbas \& Sheth 2007); 
note, however, that assembly bias effects are thought to be stronger for lower mass 
haloes (e.g., Croton et al.\ 2007; Wang et al.\ 2007), which host fainter galaxies than the SDSS can reliably probe.
The results of these studies suggest that assembly 
bias affects the formation of observed galaxies only weakly, if at all, although 
the debate is not yet resolved.

It is also commonly assumed that typical dark matter haloes are relaxed and in 
virial equilibrium.  Nevertheless, Macci\`{o} et al.\ (2007) have shown that many 
haloes are significantly unrelaxed, sometimes because they are experiencing or 
have recently experienced a merger or have not yet virialized.  In addition, 
statistical studies of galaxy groups and clusters have shown that the most 
massive galaxy in these systems is often significantly offset from the centre 
of the potential well, which may be an indication that some of the systems are 
unrelaxed (van den Bosch et al.\ 2005; Skibba et al.\ 2011).

The relation between halo formation and galaxy formation is far from clear, and 
these issues are in need of further investigation.  For example, to what extent 
halo mass determines other halo properties, and to what extent relaxed and 
unrelaxed haloes differ, remains to be seen.  The purpose of this paper is to 
examine the properties of dark matter haloes in numerical simulations, with 
principal component analysis (PCA).  This type of analysis allows us to quantify 
the correlations between the halo properties, determine which properties are the 
most important, and to what extent halo mass is the dominant property.  It also 
allows us to characterize the distinctive properties of relaxed and unrelaxed 
haloes.

This paper is organized as follows.  
We describe the dark matter halo simulations in Section~\ref{sec:sims}.  
In Section~\ref{sec:PCA}, we describe our principal component analyses 
and our choice of halo properties to analyze.  We then present the PCA 
results in Section~\ref{sec:res}.  
We further examine correlations with halo mass and relaxedness in Section~\ref{sec:relax}. 
Finally, we end with a brief discussion of 
our results and their implications in Section~\ref{sec:concl}.

\section{Numerical Simulations} 
\label{sec:sims}

\begin{table}
 \centering
 \begin{minipage}{140mm}
  \caption{N-body Simulation Parameters}
  \begin{tabular}{lccccr}
\hline  Name &  Box  size  & N  &  part. mass  &  force  soft. & Nhalo\\
& $[{\rm Mpc}]$ &  & $[h^{-1}M_{\odot}]$  & $[h^{-1}{\rm kpc}]$ & $>500$ \\
\hline\\
W5-20.1   & 20   & $250^3$ & 1.37e7  & 0.43 & 974\\ 
W5-30.1   & 30   & $300^3$ & 2.67e7  & 0.64 & 984\\ 
W5-40.1   & 40   & $250^3$ & 1.09e8  & 0.85 & 1119\\ 
W5-90.1   & 90   & $300^3$ & 7.21e8  & 1.92 & 1998\\ 
W5-90.2   & 90   & $600^3$ & 9.02e7  & 0.85 & 16161\\ 
W5-180  & 180  & $300^3$ & 5.77e9  & 3.83 & 2302\\ 
W5-300.1  & 300  & $400^3$ & 1.13e10 & 4.74 & 5845\\ 
W5-300.2  & 300  & $400^3$ & 1.13e10 & 4.74 & 5933\\ 
\hline 
\end{tabular}
\end{minipage}
\label{tab:sims}
\end{table}

Table  \ref{tab:sims}   lists  all of  the  simulations  used   in  this
work. Most of them have been already presented in Macci\`o \etal (2008) 
and Mu\~{n}oz-Cuartas et al.\ (2011).
We have run simulations for several different box sizes,  which allows us to  
probe halo masses  covering the entire
range $10^{10} \hMsun  \lta M \lta  10^{15} \hMsun$.  In
addition, in some cases we  have run multiple simulations
for the same  cosmology and box size, in order to  test for the impact
of cosmic  variance (and to increase  the final number  of dark matter
haloes).

All simulations have been performed  with \textsc{pkdgrav}, a tree code written
by Joachim Stadel and Thomas Quinn (Stadel 2001). The code uses spline
kernel softening, for which  the forces become completely Newtonian at
twice the softening length, $\epsilon$. The physical values of $\epsilon$ at $z=0$ are listed
in  Table  \ref{tab:sims}. The initial conditions  are generated with
the \textsc{grafic2} package (Bertschinger 2001).

We have set the cosmological parameters according to the fifth-year
results of the Wilkinson Microwave Anisotropy Probe mission (WMAP5,
Kogut et al 2009), namely, $\Omega_m = 0.258$, $\Omega_L = 0.742$, $n=0.963$, 
$h = 0.72$, and $\sigma_8 = 0.796$.

In  all of the  simulations,  dark  matter  haloes  are  identified  using  a
spherical overdensity (SO) algorithm. We use a time varying virial density 
contrast determined using the fitting formula presented in Mainini \etal (2003). 
We include in the halo catalogue all the haloes with more than 1000 particles
(see Macci\`o \etal 2008 for further details on our halo finding algorithm). 
Note that we include only host dark matter haloes in our analysis; halo 
substructures are not included.

\subsection{Halo Parameters}
\label{ssec:param}

For  each SO  halo in  our sample  we determine  a set  of parameters,
including the virial mass and radius, the concentration parameter, the
angular momentum, the spin  parameter and axis ratios (shape).
Below  we  briefly  describe  how  these parameters  are  defined  and
determined. A more detailed discussion can be found in Macci\`o \etal (2007, 2008).

\subsubsection{Concentration parameter}

To compute the concentration of  a halo we first determine its density
profile. The halo centre is defined  as the location of the most bound
halo particle, and  we compute the density ($\rho_i$)  in 50 spherical
shells, spaced  equally in logarithmic  radius. Errors on  the density
are  computed from  the  Poisson noise  due  to the  finite number  of
particles in  each mass shell.   The resulting density profile  is fit
with a NFW profile (Navarro \etal 1997):
\begin{equation}
\frac{\rho(r)}{\rhocrit} = \frac{\delta_{\rm c}}{(r/\rs)(1+r/\rs)^2},
\label{eq:nfw}
\end{equation}
During the  fitting procedure  we treat both  $\rs$ and  $\delta_c$ as
free  parameters.   Their values,  and  associated uncertainties,  are
obtained via a $\chi^2$  minimization procedure using the Levenberg \&
Marquardt method.  We define the r.m.s. of the fit as:
\begin{equation}
\rhorms = \frac{1}{N}\sum_i^N { (\ln \rho_i - \ln \rho_{\rm m})^2}
\label{eq:rms}
\end{equation}
where $\rho_{\rm m}$ is the fitted NFW density distribution.
Finally,   we   define  the   concentration   of   the  halo,   $\cvir
\equiv\Rvir/\rs$,  using  the  virial  radius  obtained  from  the  SO
algorithm,   and    we   define   the    error   on   $\log    c$   as
$(\sigma_{\rs}/\rs)/\ln(10)$,  where  $\sigma_{\rs}$  is  the  fitting
uncertainty on $\rs$.   

\subsubsection{Spin parameter}
\label{sec:spinpar}

The  spin  parameter is  a  dimensionless  measure  of the  amount  of
rotation of a  dark matter halo.  We use  the definition introduced by
Bullock \etal (2001):
\begin{equation}
\lam=\frac{\Jvir}{\sqrt{2}\Mvir\Vvir\Rvir}
\end{equation}
where  $\Mvir,$  $\Jvir$  and  $\Vvir$  are the  mass,  total  angular
momentum  and circular velocity  at the  virial radius,  respectively. 
See  Macci\`o \etal (2007)  for  a detailed  discussion  and  for  a comparison  of  the
different definitions of the spin parameter.
Note that the concentration and spin parameters are not well-defined in 
unrelaxed haloes; for this reason, most of our analyses in 
Section~\ref{sec:res} involve only relaxed haloes.

\subsubsection{Shape parameter}

Determining the shape of a three-dimensional distribution of particles
is a  non-trivial task (e.g.,  Jing \& Suto 2002).   Following Allgood
\etal (2006), we  determine the shapes of our  haloes starting from the
inertia tensor.  As  a first step, we compute the  halo's $3 \times 3$
inertia tensor using all the particles within the virial radius.  Next, 
we diagonalize the inertia tensor and rotate the particle distribution
according  to the  eigenvectors.   In this  new frame  (in  which the
moment  of inertia  tensor  is diagonal)  the  ratios $a_3/a_1$  and
$a_3/a_2$ (where $a_1 \geq a_2 \geq a_3$) are given by:
\begin{equation}
{a_3 \over a_1} = \sqrt{ { \sum m_i z_i^2} \over \sum { m_i x_i^2}}\\
{a_3 \over a_2} = \sqrt{ { \sum m_i z_i^2} \over \sum { m_i y_i^2}}.
\end{equation}

Next we again compute the inertia tensor, but this time only using the
particles  inside the  ellipsoid defined  by $a_1$,  $a_2$, and  $a_3$.
When deforming the ellipsoidal volume of the halo, we keep the longest
axis  ($a_1$) equal  to the  original radius  of the  spherical volume
($\Rvir$).  We  iterate this procedure  until we converge to  a stable
set of axis ratios. 

Since dark matter haloes tend to be more prolate on average (e.g., Kasun \& Evrard 2005; 
Bett et al.\ 2007), a useful 
parameter that describes the shape of the halo is $q\equiv (a_2+a_3)/2a_1$, with 
the limiting cases being a sphere ($q=1$) and a needle ($q=0$). 
A related parameter is the triaxiality $\tau\equiv (a_1^2-a_2^2)/(a_1^2-a_3^2)$, 
which can also be used to quantify the prolateness of haloes, but we find that 
it yields nearly identical results as the shape parameter, so we choose to use 
only $q$ in Section~\ref{sec:res}.

Note that for haloes that are not spherical, the NFW fit to the density profile could 
be affected, potentially inducing an `artificial' correlation between halo shape 
and the r.m.s. of the density profile fit.  Nevertheless, although triaxial models 
are an improvement over spherical profile fits, Jing \& Suto (2002) have shown that 
the differences between such fitted profiles are fairly small, and in any case, Macci\`o 
et al.\ (2007) have shown that there is no significant correlation between halo shape 
and the r.m.s. of the spherical profile fit.

\subsubsection{Position angle $\Phi_L$}

We also consider the angle $\Phi_L$ between the major axis of the halo 
($a_1$) and the angular momentum vector $L$. 
The angle $\Phi_L$ can be thought of as a proxy for the position 
of a possible stellar disk (and for the orientation of satellite galaxies) 
within the dark matter halo, assuming that the 
angular momentum of the disk and the one of the dark matter particles 
will be aligned (e.g., Sharma \& Steinmetz 2005; Zentner et al.\ 2005b; Heller et al.\ 2007; Hahn et al.\ 2010). 
Nevertheless, it is difficult for simulations to reproduce the 
angular momentum distributions of observed disk galaxies 
(van den Bosch 2002; Sharma \& Steinmetz 2005).  More recently, Heller et al.\ (2007) 
claim to reproduce many of the properties of observed galactic disks, 
and the angular momentum of their simulated disks follows that of the dark matter and does not decrease too much while the disks form.
Agertz et al.\ (2010) also can reproduce the angular momentum content of disk galaxies 
by assuming a lower star formation efficiency than other simulations and weaker feedback energy from supernovae. 
In contrast, Governato et al.\ (2010) argue that strong outflows from supernovae are 
necessary to reproduce the angular momentum distribution of dwarf galaxies, and help to yield shallow dark matter profiles. 
%

We note that the shape parameter $q$ and $\Phi_L$ are not estimated from all 
of the particles in a halo, but rather from the halo's best-fit ellipsoid. 
Nevertheless, for the vast majority of the haloes, these parameters are computed 
using $>80\%$ of the particles, so we expect that this does not affect or bias 
our results significantly.
In addition, observational estimates of the position angles of 
galaxies also have large uncertainties (e.g., Barnes \& Sellwood 2003). 
%
In any case, although for a spherical halo $\Phi_L$ will be poorly defined, 
this does not result in systematic uncertainties in our analysis or in 
artificial correlations between $\Phi_L$ and halo shape.

\subsubsection{Environment parameter $\Delta_8$}

Finally, in some of our analyses we also consider the overdensity ($\Delta_8$), 
computed within a sphere of radius $R=8 \mpch$ centred on the halo centre.
Such an overdensity within a fixed aperture allows us to quantify the 
large-scale environment of haloes. 
Although most haloes are smaller than 2 Mpc/$h$ in size (e.g., Navarro, Frenk \& White 1997), 
it is useful to focus on larger scales, in order to more clearly encompass 
nonlinear scales within the aperture (e.g., Abbas \& Sheth 2007).  

We use the 8 Mpc/$h$ overdensity, $\Delta_8$, in our analyses in Section~\ref{sec:res}. 
In the appendix, we show similar results for smaller scale overdensities,  
$\Delta_2$ and $\Delta_4$.

\subsection{Relaxed and Unrelaxed Haloes}
\label{sub:relax}

Our halo  finder, and  halo finders in  general, do  not distinguish
between relaxed and  unrelaxed haloes. 
(although there are recent exceptions, using halo phase-space densities: e.g., 
Zemp et al.\ 2009; Behroozi et al., in prep.). 
There are many  reasons why we
might want  to remove unrelaxed haloes.  First  and foremost, unrelaxed
haloes   often  have   poorly   defined  centers,   which  makes   the
determination  of   a  radial  density  profile,  and   hence  of  the
concentration parameter,  an ill-defined problem.   Moreover unrelaxed
haloes  often have  shapes that  are  not adequately  described by  an
ellipsoid, making our shape parameters ill-defined as well.

Following Macci\`o \etal (2007, 2008), we decide to use a combination of
two different parameters $\rhorms$ and $\xoff$  to determine the dynamical status 
of a given dark matter halo. The first quantity $\rhorms$ is defined as
the r.m.s.  of the  NFW fit to the density profile (performed to compute $c_{vir}$).
While  it is  true  that  $\rhorms$ is  typically  high for  unrelaxed
haloes,  haloes  with  relatively  few  particles  also  have  a  high
$\rhorms$  (due to  Poisson noise)  even when  they are  relaxed; furthermore, 
since the spherical averaging used to compute the density profiles has a  
smoothing effect, not all unrelaxed  haloes have a high $\rhorms$. 

In order to circumvent these problems, we combine the value of $\rhorms$ with 
the $\xoff$ parameter, defined as the  distance between the most bound  particle 
(used as the center for the density profile) and the center of mass of the halo, in
units of the virial radius. 
This offset is a measure for the extent to  which the  halo is relaxed:  
relaxed haloes  in equilibrium will have a smooth,  radially symmetric density distribution, and thus
an offset that is virtually  equal to zero.  Unrelaxed haloes, such as
those that have  only recently experienced a major  merger, are likely
to  reveal  a  strongly  asymmetric  mass  distribution,  and  thus  a
relatively large  $\xoff$. 

Although some unrelaxed haloes  may have a small $\xoff$, the
advantage  of this  parameter  over, for  example,  the actual  virial
ratio, $2T/V$, as a function of radius, 
is that the former is trivial to evaluate.
In addition, the virial ratio can be quite noisy (e.g., Macci\`o, Murante \& Bonometto 
2003), while the substructure fraction, another proxy for virialization, depends significantly 
on the simulation's resolution (Macci\`o et al. 2007). 

\begin{figure}
\includegraphics[width=0.5\textwidth]{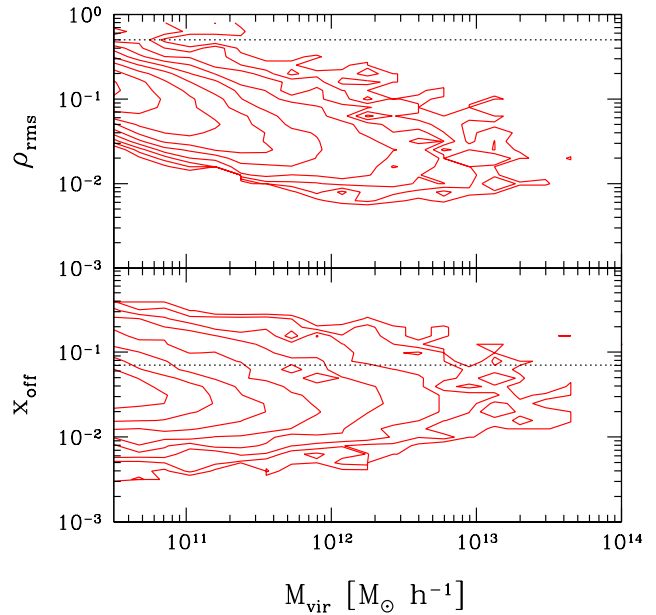}
\caption{Correlation between the relaxedness parameters ($\xoff$ and $\rhorms$ and the virial mass $\Mvir$.
The two dotted lines represents the limit values used to define `good' haloes.}
\label{fig:dens}
\end{figure}

Figure \ref{fig:dens} shows the correlation between $\xoff$ (lower  panel), 
$\rhorms$ (upper panel), and the virial mass of the halo for the W5-90.2 box. As already noted in 
Macci\`o \etal (2007), there is a correlation between the $\Mvir$ (or $\Nvir$)
and $\rhorms$, which simply reflects the fact that for low halo masses the 
density profiles are more noisy. 
In most of our analyses in Section~\ref{sec:res}, we use a dark matter particle threshold as well as a mass criterion in order to minimize any effect of this correlation on our results.
Finally, $\xoff$ and $\Mvir$ are 
extremely weakly correlated, if at all, and we find no correlation between  $\xoff$ and $\rhorms$.

Following Macci\`o \etal (2007), we split our halo sample into unrelaxed and relaxed haloes.
The latter are defined as the haloes with $\rhorms  < 0.5$ and $\xoff < 0.07$.  About
70\% of  the haloes in our  sample qualify as relaxed  haloes. In what
follows, we will  present results for two different  samples of haloes:
`all', which includes all haloes  with $\Nvir > 1000$, and `good' or `relaxed', which
is the corresponding subsample of relaxed haloes.

\section{Principal Component Analysis} 
\label{sec:PCA}

\subsection{Introduction to PCA}
\label{sec:PCAintro}

Principal component analysis (PCA) is a type of multivariate (multidimensional) 
statistics. The goal of such an analysis is to find a suitable representation of 
multivariate data, where representation here means that we somehow transform the 
data so that its essential structure is made more visible or accessible.  PCA 
belongs to the area of unsupervised learning, because the representation must be 
learned from the data itself without any external input from a supervising 
``teacher''. 

PCA is a powerful technique that has been used in classification and dimensional 
reduction of large data sets, in order to decrease the complexity of the 
analysis.  Each member of a data set (a dark matter halo, in our case) is 
defined by an ``information vector'', an $n$-tuple of numbers (the halo's 
relevant properties, in our case).  PCA consists of de-correlating sets of 
vectors by performing rotations in the $n$-dimensional parameter space.  The 
final result is a diagonal covariance matrix, with the eigenvalues representing 
the amount of information (in the sense of variance) stored in each eigenvector, 
which is called a principal component. 

The principal components can be considered a set of basis vectors that optimally 
filter the information hidden in the data.  If most of the variance in the 
original data can be accounted for by just a few of the components, we will have 
found a simpler description of the original data set.  By convention, the first 
principal component corresponds to the largest eigenvalue, the second principal 
component corresponds to the second largest eigenvalue, and so on.  A 
consequence of this definition is that the first principal component (PC) is a 
minimum distance fit to a line in the space of the original variables.  The 
second PC is then a minimum distance fit to a line in the plane perpendicular to 
the first PC, and so on.  Therefore, the properties that dominate the first PC 
are the properties that contain the largest amount of information and to some 
extent determine the values of the other properties.  Typically the eigenvalue 
for a PC must be $\geq 1$ to be significant; less significant PCs tend to be 
dominated by noise (e.g., Connolly \& Szalay 1999). 
In any case, the amount of variance that a PC accounts for is 
more meaningful than the precise value of its eigenvalue. 
If the data have high signal-to-noise, and if the data are not very incomplete, 
then the amount of variance of the first few PCs can be associated with `information' 
contained in the data. 

In astronomy, PCA applications in studies of multivariate distributions have 
been discussed in detail (Efstathiou \& Fall 1984; Murtagh \& Heck 1987).  PCA 
methods have been used to study stellar, galaxy, and quasar spectra (e.g., Connolly et al.\ 
1995; Bailer-Jones et al.\ 1998; Madgwick et al.\ 2003; Yip et al.\  2004; Ferreras et al.\ 2006; 
Rogers et al.\ 2007; Chen et al.\ 2009; McGurk et al.\ 2010; Boroson \& Lauer 2010), 
galaxy properties (e.g., Conselice 2006; Scarlata et al.\ 2007), and stellar populations 
and the Fundamental Plane (e.g., Faber 1973; Eisenstein et al.\ 2003; Woo et al.\ 2008). 
In a recent study, Bonoli \& Pen (2009) analyzed the 
stochasticity of dark matter haloes, which quantifies the scatter in the correlation 
between halo and galaxy density fields.  

PCA methods have not yet been exploited to analyze the properties of dark matter 
haloes.  Considering the increasing size and resolution of numerical simulations, 
PCA is a useful tool to determine the relative importance of halo properties, 
and to determine how fundamental the masses of haloes are.  The primary goal of 
this paper is to use PCAs to analyze a variety of halo properties, including those 
introduced in Section~\ref{sec:sims}.

\subsection{Applying PCAs to DM Haloes}

Before we perform our principal component analyses on our simulation data, we 
subtract the mean from each halo property and normalize by the standard 
deviation.  It follows that the distributions of the properties may be 
important: for example, choices about whether to take the logarithm of a 
property or whether to include objects whose properties are outliers in the 
distributions may be important.
It is particularly useful to take the logarithm of halo mass, for example, because its range 
spans many orders of magnitude, and the spin parameter, because this makes its 
distribution more Gaussian; for consistency, however, we take the logarithms 
of all of the halo properties in our PCA.
We also choose to 
exclude the extreme outliers ($>5\sigma$) in the distributions of properties.  In addition, 
some properties are so strongly correlated that one property almost 
completely determines another (such as the virial mass, radius and velocity, or 
spin and angular momentum, or shape and triaxiality), 
and in such a case we include only what we believe is the more 
important property.  By reducing the number of parameters, 
this tends to clarify the results of the PCA.  We have 
performed numerous tests and have verified that these choices do not 
significantly affect our results.

We perform PCAs of each of the simulations described in Section~\ref{sec:sims}. 
We analyze all of the haloes in each box, as well as the `relaxed' haloes, with low values of $\rho_\mathrm{rms}$ (the r.m.s. of the NFW fit to the density profile) and $x_\mathrm{off}$ (the offset between the most bound particle and the center of mass).

\begin{figure}
\includegraphics[width=0.5\textwidth]{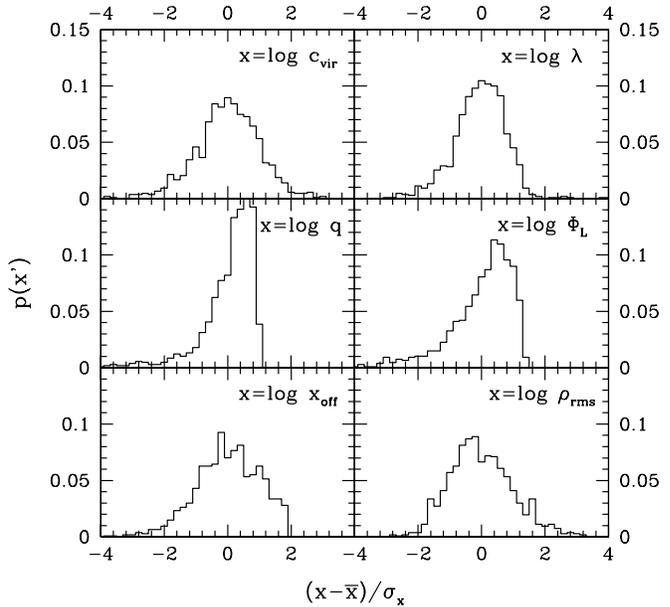}
\caption{
         Distributions of halo properties at fixed mass (${\rm log}~(M/h^{-1}M_\odot)\sim12$), 
         for haloes with more than 1000 particles in the high resolution 90~Mpc box. 
         For a given property $x$, we define the mean-subtracted and standard 
         deviation-normalized quantity as $x'\equiv(x-{\bar x})/\sigma_x$.
        }
\label{dists}
\end{figure}

The probability distributions of the halo properties are shown in 
Figure~\ref{dists}, for haloes in the 90~Mpc box, at fixed halo mass (see 
Section~\ref{sec:fixedM}).  By construction, PCA only finds independent normal 
processes, and works most effectively when the data set is jointly normally 
distributed.  Our halo parameters are chosen to be approximately independent, and 
as can be seen in the figure, their distributions are approximately Gaussian.  The 
shape parameter $q$ and the position angle $\Phi_L$ have slightly non-Gaussian 
distributions, with a tail of low values, but this is not 
sufficient to significantly affect the PCA results. 
\begin{table}
 \begin{center}
 \begin{tabular}[h!]{ l | c c c }
   property & median & mean & std dev \\
   \hline
   ${\rm log}\,c_{\rm vir}$ & 1.043 & 1.039 & 0.107 \\
   ${\rm log}\,\lambda$ & -1.498 & -1.513 & 0.252 \\
   ${\rm log}\,q$ & -0.101 & -0.126 & 0.094 \\
   ${\rm log}\,\Phi_L$ & 0.207 & 0.147 & 0.249 \\
   ${\rm log}\,x_{\rm off}$ & -1.614 & -1.622 & 0.248 \\
   ${\rm log}\,\rho_{\rm rms}$ & -1.439 & -1.413 & 0.259 \\
   \hline
 \end{tabular}
 \end{center}
 \caption{
          Median, mean, and standard deviation of halo properties at fixed mass in the 
          90~Mpc box, whose distributions are shown in Figure~\ref{dists}.
         }
 \label{propstats}
\end{table}

\begin{figure}
\includegraphics[width=0.5\textwidth]{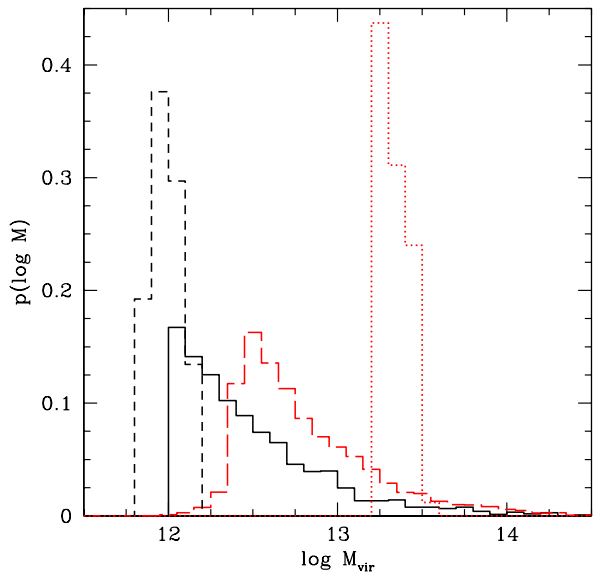} 
\caption{
         Log halo mass distributions for ${\rm log}~(M/h^{-1}M_\odot)\geq12$, 90~Mpc box (solid black histogram); ${\rm log}~(M/h^{-1}M_\odot)\geq12$, 180~Mpc box (long-dashed red histogram); relaxed haloes with $11.85\leq \mathrm{log}~(M/h^{-1}M_\odot)<12.15$, 90~Mpc box (short-dashed black histogram); relaxed haloes with $13.2\leq \mathrm{log}~(M/h^{-1}M_\odot)<13.5$, 180~Mpc box (dotted red histogram).
        }
\label{Mdists}
\end{figure}

Lastly, we show the log halo mass distributions in Figure~\ref{Mdists}, for some of the analyses in Section~\ref{sec:res}.  (In particular, these are the distributions for the PCAs whose results are shown in Tables~\ref{all90}, \ref{all180}, \ref{good90M}, and \ref{good180M}.) 
We will perform PCAs with halo mass thresholds (Section~\ref{sec:Mmin}), followed by PCAs at fixed halo mass (Section~\ref{sec:fixedM}), and the mass distributions of some of these halo catalogues are shown in the figure. 
When a halo mass threshold is used, the mass distributions are highly non-Gaussian; nonetheless, the effect of this on the PCA results is minimal, as we discuss in the next section. 

\section{Results}
\label{sec:res}

For the principal component analyses, we focus on eight halo properties: 
mass, concentration, spin, shape, $\Phi_L$, $x_\mathrm{off}$, $\rho_\mathrm{rms}$, 
and overdensity.  Firstly, in Section~\ref{sec:Mmin}, we perform PCA of the haloes in our simulations using a halo mass threshold. 
Then in Section~\ref{sec:fixedM}, we perform PCA at a fixed halo mass.
Lastly, we briefly discuss the lack of a correlation between $\Phi_L$ and other 
halo parameters in Section~\ref{sec:phil}.

\subsection{PCAs with a halo mass threshold}
\label{sec:Mmin}

The results of our principal component analyses of all of the haloes in the 
90 and 180~Mpc boxes with ${\rm log}~(M/h^{-1}M_\odot)\geq12$ are shown in Tables~\ref{all90} and \ref{all180}; 
the results for the subset of relaxed haloes are shown in Tables~\ref{good90} and \ref{good180}. 
Similar PCA results for smaller and larger simulation boxes are shown in the appendix. 
Note that because of our resolution constraint, in the 180~Mpc box, 
most of the haloes with ${\rm log}~(M/h^{-1}M_\odot)\geq12$ are more massive than $10^{12.3}h^{-1}M_\odot$ (see Fig.~\ref{Mdists}).

We consider the following seven halo properties, of which we use the logarithms: 
the mass ($M_\mathrm{vir}$), the concentration ($c_\mathrm{vir}$), the spin ($\lambda$), 
the shape ($q$), the angle between the major axis and the total angular momentum ($\Phi_L$), 
$\rho_\mathrm{rms}$, and $x_\mathrm{off}$. 
We have performed tests of the uncertainties of the PCAs with bootstrap techniques and have 
found that our errors are extremely small; hence, we have omitted them from the 
tables, for clarity.  In addition, our results are not significantly dependent on 
the selection criteria (i.e., the choice of the halo mass threshold, and for the PCAs 
later in this section, the particle limit and the $x_\mathrm{off}$ and $\rho_\mathrm{rms}$ criteria).

One striking result is the \textit{lack} of a single or pair of dominant principal components. 
The first two principal components usually account for only about 40-50\% of the variance. 
In contrast, in PCAs of galaxy spectra, the first one or two PCs usually dominate 
(e.g., Connolly et al.\ 1995; Madgwick et al.\ 2003). 
Recall that, as stated in Section~\ref{sec:PCAintro}, the later PCs 
(with lower eigenvalues) tend to be dominated by noise.

For dark matter haloes, no parameter or combination of parameters strongly determine 
the other parameters.  (This is partly by construction, because for closely related parameters, 
like mass and radius or shape and triaxiality, we have included only one of the parameters.)
In other words, although halo properties are correlated with each other, 
we find no clearly fundamental correlation (nor a fundamental `plane'). 

\begin{figure}
\includegraphics[width=0.5\textwidth]{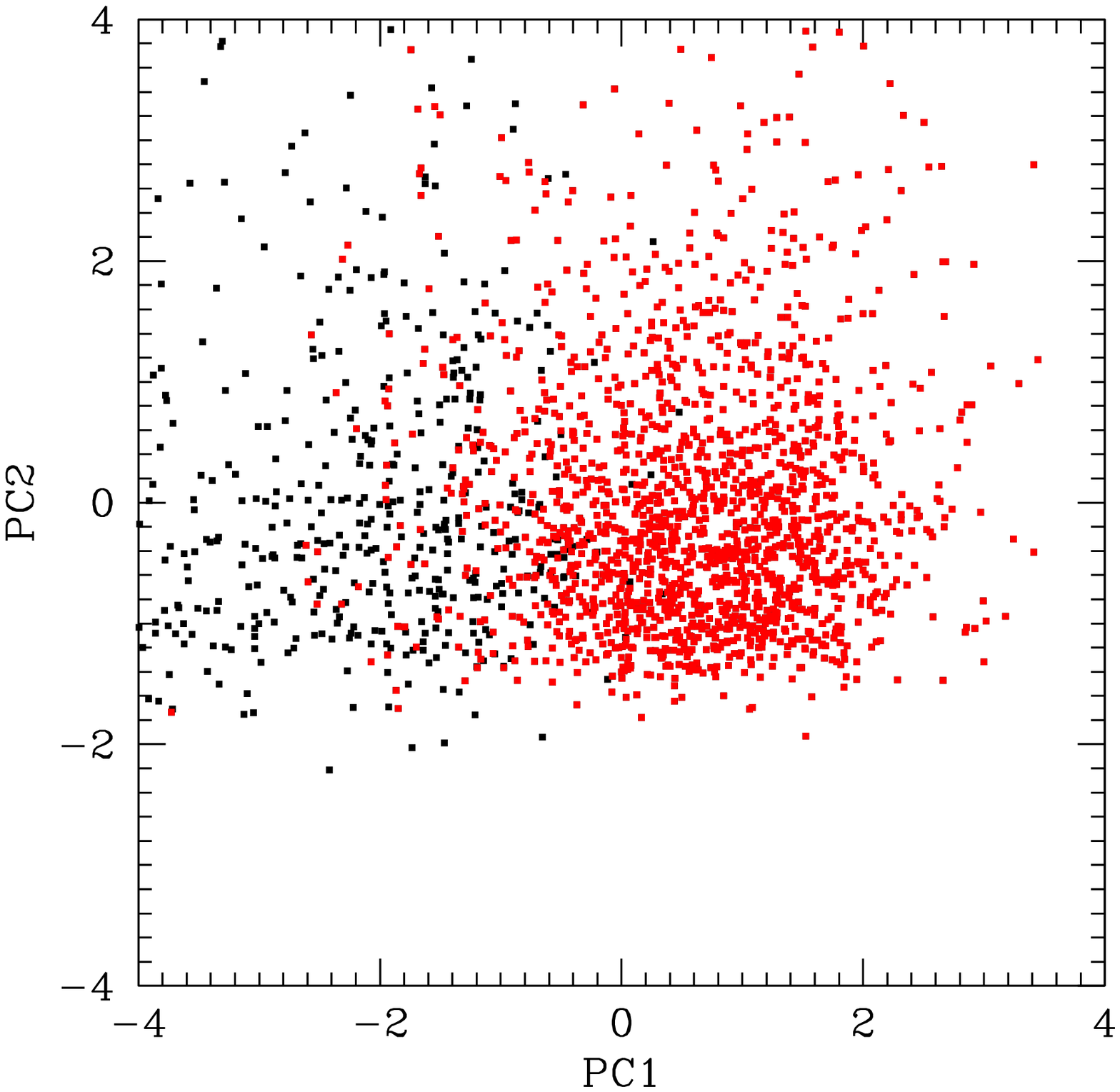}
\caption{
         First two principal components plotted against each other, for all haloes 
         with ${\rm log}~(M/h^{-1}M_\odot)\geq12$ in the 90~Mpc box. 
         We are using the results from the first two columns of Table~\ref{all90}: 
         for example, for a given halo, 
         PC1$=-0.103\,(\mathrm{log}\,M_\mathrm{vir})^{\prime}+0.555\,(\mathrm{log}\,c_\mathrm{vir})^{\prime}-0.369\,(\mathrm{log}\,\lambda)^{\prime}-0.094\,(\mathrm{log}\,q)^{\prime}-0.001\,(\mathrm{log}\,\Phi_L)^{\prime}-0.547\,(\mathrm{log}\,x_\mathrm{off})^{\prime}-0.488\,(\mathrm{log}\,\rho_\mathrm{rms})^{\prime}$, 
         where the halo parameters are mean-subtracted and normalized by the standard deviation. 
         The red points are the subset of `good' (i.e., relaxed) haloes.}
\label{fig:PC12}
\end{figure}

This is also evident in Figure~\ref{fig:PC12}, in which the first two PCs of all 
haloes in the 90~Mpc box (Table~\ref{all90}) are shown.  
Some structure is apparent in the figure, with relaxed and unrelaxed haloes exhibiting different distributions. 
The figure illustrates that the relaxed haloes have larger values of PC1 (which is anti-correlated with the relaxedness parameters), and the fact that most haloes have $\mathrm{PC}2>-2$ is simply due to the halo mass threshold.

The PCA results of the different simulations are similar, but not equivalent, 
most likely because of the different ranges of halo masses. 
In any case, in the majority of our simulation boxes, combinations of 
$M_\mathrm{vir}$, $c_\mathrm{vir}$, $x_\mathrm{off}$, and $\rho_\mathrm{rms}$ 
dominate the first two principal components. 
Halo mass and concentration, on the one hand, and halo relaxedness, on the other, 
can explain at least half of the variance. 
Note, however, that halo mass is never significant on the first PC. 
$x_\mathrm{off}$ and $\rho_\mathrm{rms}$ 
indicate the relaxedness of haloes, and for unrelaxed haloes, the concentration is not 
well-defined.  Therefore, \textit{we argue that the relaxedness of haloes is at least as 
important as their mass and concentration} for determining halo properties. 
This is the primary result of our paper.

\begin{table}
 \begin{center}
 \begin{tabular}[h!]{ l | c c c c }
   property & PC 1 & PC 2 & PC 3 & PC 4 \\
   \hline
   ${\rm log}\,M_{\rm vir}$ & 0.103 & {\bf -0.761} & 0.148 & {\bf 0.500} \\
   ${\rm log}\,c_{\rm vir}$ & {\bf -0.555} & 0.156 & -0.015 & -0.134 \\
   ${\rm log}\,\lambda$ & 0.369 & 0.009 & 0.062 & -0.333 \\
   ${\rm log}\,q$ & 0.094 & {\bf 0.579} & 0.143 & {\bf 0.759} \\
   ${\rm log}\,\Phi_L$ & 0.001 & -0.059 & {\bf -0.971} & 0.181 \\
   ${\rm log}\,x_{\rm off}$ & {\bf 0.547} & -0.018 & -0.012 & -0.041 \\
   ${\rm log}\,\rho_{\rm rms}$ & {\bf 0.488} & 0.240 & -0.107 & -0.105 \\
   \hline
   eigenvalues & 2.56 & 1.18 & 1.01 & 0.90 \\
   \% of variance & 36.5\% & 16.9\% & 14.4\% & 12.8\% \\
  \end{tabular}
 \end{center}
 \caption{All haloes with ${\rm log}~(M/h^{-1}M_\odot)\geq12$, 90~Mpc box, high resolution. $N$=2183.}
 \label{all90}
 \begin{center}
 \begin{tabular}[h!]{ l | c c c c }
   property & PC 1 & PC 2 & PC 3 & PC 4 \\
   \hline
   ${\rm log}\,M_{\rm vir}$ & 0.163 & {\bf -0.762} & 0.064 & 0.166 \\
   ${\rm log}\,c_{\rm vir}$ & {\bf -0.604} & 0.231 & 0.006 & -0.114 \\
   ${\rm log}\,\lambda$ & 0.392 & 0.102 & 0.128 & -0.221 \\
   ${\rm log}\,q$ & 0.085 & 0.307 & 0.304 & {\bf 0.887} \\
   ${\rm log}\,\Phi_L$ & 0.033 & -0.025 & {\bf -0.928} & 0.300 \\
   ${\rm log}\,x_{\rm off}$ & {\bf 0.543} & -0.001 & 0.021 & -0.047 \\
   ${\rm log}\,\rho_{\rm rms}$ & 0.389 & {\bf 0.511} & -0.162 & -0.176 \\
   \hline
   eigenvalues & 1.84 & 1.35 & 1.01 & 0.96 \\
   \% of variance & 26.3\% & 19.2\% & 14.5\% & 13.7\% \\
  \end{tabular}
 \end{center}
 \caption{`Good' (relaxed) haloes with ${\rm log}~(M/h^{-1}M_\odot)\geq12$, 90~Mpc box, high resolution. $N$=1701.}
 \label{good90}
\end{table}
\begin{table}
 \begin{center}
 \begin{tabular}[t!]{ l | c c c c }
   property & PC 1 & PC 2 & PC 3 & PC 4 \\
   \hline
   ${\rm log}\,M_{\rm vir}$ & {\bf 0.480} & {\bf 0.467} & 0.025 & -0.078 \\
   ${\rm log}\,c_{\rm vir}$ & 0.221 & {\bf -0.639} & -0.022 & 0.007 \\
   ${\rm log}\,\lambda$ & -0.320 & {\bf 0.417} & -0.033 & 0.046 \\
   ${\rm log}\,q$ & -0.218 & -0.023 & 0.263 & {\bf -0.936} \\
   ${\rm log}\,\Phi_L$ & 0.013 & 0.001 & {\bf 0.963} & 0.268 \\
   ${\rm log}\,x_{\rm off}$ & {\bf -0.539} & 0.257 & -0.040 & 0.167 \\
   ${\rm log}\,\rho_{\rm rms}$ & {\bf -0.530} & -0.358 & -0.011 & 0.126 \\
   \hline
   eigenvalues & 1.94 & 1.64 & 1.00 & 0.95 \\
   \% of variance & 27.7\% & 23.5\% & 14.3\% & 13.6\% \\
  \end{tabular}
 \end{center}
 \caption{All haloes with ${\rm log}~(M/h^{-1}M_\odot)\geq12$, 180~Mpc box. $N$=6391.}
 \label{all180}
 \begin{center}
 \begin{tabular}[t!]{ l | c c c c }
   property & PC 1 & PC 2 & PC 3 & PC 4 \\
   \hline
   ${\rm log}\,M_{\rm vir}$ & {\bf -0.618} & -0.098 & 0.030 & 0.011 \\
   ${\rm log}\,c_{\rm vir}$ & {\bf 0.443} & {\bf -0.469} & -0.003 & 0.044 \\
   ${\rm log}\,\lambda$ & -0.062 & {\bf 0.664} & -0.044 & -0.012 \\
   ${\rm log}\,q$ & 0.113 & 0.286 & 0.404 & {\bf 0.802} \\
   ${\rm log}\,\Phi_L$ & 0.010 & -0.004 & {\bf 0.899} & -0.435 \\
   ${\rm log}\,x_{\rm off}$ & 0.236 & {\bf 0.494} & -0.164 & -0.406 \\
   ${\rm log}\,\rho_{\rm rms}$ & {\bf 0.591} & 0.066 & 0.002 & -0.008 \\
   \hline
   eigenvalues & 2.06 & 1.25 & 1.01 & 0.98 \\
   \% of variance & 29.4\% & 17.8\% & 14.4\% & 14.0\% \\
  \end{tabular}
 \end{center}
 \caption{`Good' (relaxed) haloes with ${\rm log}~(M/h^{-1}M_\odot)\geq12$, 180~Mpc box. $N$=3757.}
 \label{good180}
\end{table}
 
When halo parameters appear together on a PC, this can indicate a correlation between those parameters.  For example, when a PC is dominated by mass and concentration, they usually have opposite signs, consistent with the fact that these halo properties are anti-correlated (e.g., Bullock et al.\ 2001).  Correlated halo properties do not \textit{necessarily} dominate the same PCs, however, and properties dominating a PC are not necessarily correlated, although this is often the case.  
If two correlated properties have substantial scatter between them, or if their correlation is due to more fundamental correlations with a third parameter, or if two properties have strongly correlated errors, the interpretation of PCA results may be more complicated.  With our choice of halo properties and selection criteria, we have attempted to minimize these complications.

Finally, note the similarities and differences between the results for 
all haloes and for just the relaxed haloes.  Most of the same halo parameters 
dominate the first few principle components. 
In addition, the PCA results are mostly similar in the different resolution 
simulation boxes, indicating that our results are not very sensitive to resolution, 
though the significance of the spin parameter varies slightly. 
Although we find that the PCA results presented here are statistically stable, in some cases the first PC of one simulation box resembles the second PC of another, and vice versa, suggesting that it is the combination of the first two PCs that is important. 
In any case, throughout our analyses, halo mass and concentration on the one hand, 
and relaxedness parameters (quantified by $x_\mathrm{off}$ and $\rho_\mathrm{rms}$) 
on the other, continue to dominate the first two PCs. 

We also note that the halo mass distributions are non-Gaussian when a mass threshold is used (Fig.~\ref{Mdists}), and this could potentially affect the PCA results. 
We have tested this effect by performing PCAs with subsamples that 
have Gaussian mass distributions, and by PCAs with the whole samples but with the mass 
distributions rescaled (forcing them to be Gaussian, with the same rank order), and we find that 
the results are virtually the same, differing from Tables~\ref{all90}-\ref{good180} by $<1\%$.

Nonetheless, some halo parameters are accurately determined only for relaxed haloes 
(Macci\`{o} et al.\ 2007), such that the very unrelaxed haloes introduce noise in the other parameters and could relegate them to the later PCs, or could introduce artificial correlations with the relaxedness parameters. 
For these reasons, 
for the rest of this paper, we focus on PCA results based on the subset of relaxed haloes.


\subsection{PCAs at fixed halo mass}
\label{sec:fixedM}

As shown in Figure~\ref{dists}, at fixed halo mass, the distributions of the logarithm of 
most halo properties are approximately Gaussian (i.e., $p(\mathrm{log}\,\lambda|M)$, 
$p(\mathrm{log}\,c|M)$, 
etc., can be fit fairly well by Gaussian distributions). 
It therefore makes sense to perform PCAs of these variates at fixed mass.
Constraints from these PCAs could be useful for halo model analyses (e.g., of 
galaxy clustering, weak lensing, galaxy groups and clusters), and perhaps for 
semi-analytic models as well, because they reduce the complexity of such analyses 
to a Gaussian Mixture model.  Gaussian Mixture models have a well developed 
statistical framework and have been applied to a variety of astrophysical 
studies (e.g., Kelly \& McKay 2004; Hao et al.\ 2009; Skibba \& Sheth 2009).  

We first perform PCAs similar to those in Section~\ref{sec:Mmin}, 
but now at fixed halo mass, using the other six halo parameters.
The results of these PCAs are shown in Table~\ref{good90M} for haloes with 
$11.85\leq \mathrm{log}~(M/h^{-1}M_\odot)<12.15$ in the 90~Mpc box, and in Table~\ref{good180M} 
for haloes with $13.2\leq \mathrm{log}~(M/h^{-1}M_\odot)<13.5$ in the 180~Mpc box. 
Only haloes with more than 1000 particles are included. 
\begin{table}
 \begin{center}
 \begin{tabular}[h!]{ l | c c c c }
   property & PC 1 & PC 2 & PC 3 & PC 4 \\
   \hline
   ${\rm log}\,c_{\rm vir}$ & {\bf -0.601} & -0.076 & 0.044 & -0.011 \\
   ${\rm log}\,\lambda$ & {\bf 0.441} & -0.089 & {\bf 0.544} & -0.048 \\
   ${\rm log}\,q$ & 0.167 & {\bf -0.591} & {\bf -0.427} & {\bf -0.654} \\
   ${\rm log}\,\Phi_L$ & -0.023 & {\bf 0.772} & -0.113 & {\bf -0.605} \\
   ${\rm log}\,x_{\rm off}$ & {\bf 0.552} & 0.088 & 0.158 & -0.071 \\
   ${\rm log}\,\rho_{\rm rms}$ & 0.334 & 0.183 & {\bf -0.694} & {\bf 0.446} \\
   \hline
   eigenvalues & 1.79 & 1.06 & 0.97 & 0.93 \\
   \% of variance & 29.8\% & 17.6\% & 16.2\% & 15.4\% \\
  \end{tabular}
 \end{center}
 \caption{`Good' (relaxed) haloes with ${\rm log}~(M/h^{-1}M_\odot)\sim12$ and more than 1000 particles, 90~Mpc box, high resolution. $N$=1056.}
 \label{good90M}
 \begin{center}
 \begin{tabular}[h!]{ l | c c c c }
   property & PC 1 & PC 2 & PC 3 & PC 4 \\
   \hline
   ${\rm log}\,c_{\rm vir}$ & {\bf -0.643} & 0.107 & 0.015 & 0.000 \\
   ${\rm log}\,\lambda$ & {\bf 0.496} & 0.141 & 0.339 & 0.120 \\
   ${\rm log}\,q$ & 0.181 & {\bf -0.567} & -0.389 & {\bf -0.642} \\
   ${\rm log}\,\Phi_L$ & 0.052 & {\bf -0.588} & {\bf 0.715} & 0.058 \\
   ${\rm log}\,x_{\rm off}$ & {\bf 0.520} & 0.058 & -0.353 & 0.344 \\
   ${\rm log}\,\rho_{\rm rms}$ & -0.186 & {\bf -0.546} & -0.313 & {\bf 0.672} \\
   \hline
   eigenvalues & 1.55 & 1.13 & 0.97 & 0.93 \\
   \% of variance & 25.8\% & 18.8\% & 16.2\% & 15.6\% \\
  \end{tabular}
 \end{center}
 \caption{`Good' (relaxed) haloes with ${\rm log}~(M/h^{-1}M_\odot)\sim13.35$ and more than 1000 particles, 180~Mpc box. $N$=416.}
 \label{good180M}
\end{table}

The results are similar to those of the previous section, with concentration 
and $x_\mathrm{off}$ dominating the first principal component, although the 
spin parameter is now important on this PC as well.
In addition, $\Phi_L$ and the shape parameter dominate the second PC.
Halo model analyses often involve constructing mock galaxy catalogues, and the 
models are often based on the clustering and abundance of haloes as a function of 
their masses; at fixed mass, the models use halo concentrations and density profiles. 
Perhaps the models could also use halo spin distributions, to more accurately 
describe galaxy velocities, as well as $x_\mathrm{off}$ distributions, to 
describe central galaxies offset from the halo centers.

Next, we perform PCAs with an additional parameter, the 
overdensity $\Delta_8$ within a sphere of radius 8 Mpc, which is an 
indicator of a halo's local environment (since the largest haloes have virial radii of 
$\sim2\,\mathrm{Mpc}/h$). 
At a fixed halo mass, there may be a range of environments: some haloes may be 
embedded within large structures or filaments or surrounded by infalling haloes, 
while some may be relatively isolated (e.g., Colberg et al.\ 1999).
Using a fixed halo mass also ensures that the haloes are of similar size and are 
much smaller than the density aperture.
We have restricted this analysis to haloes with more than 1000 particles and with 
$13.15<\mathrm{log}~(M/h^{-1}M_\odot)<13.45$ in the 180~Mpc box and 
$13.45<\mathrm{log}~(M/h^{-1}M_\odot)<13.75$ in one of the 300~Mpc boxes.  It is useful to analyze the 
haloes in a large simulation box, in which cosmic variance is relatively unimportant. 
\begin{table}
 \begin{center}
 \begin{tabular}[h!]{ l | c c c c }
   property & PC 1 & PC 2 & PC 3 & PC 4 \\
   \hline
   ${\rm log}\,\Delta_8$ & -0.252 & -0.103 & {\bf 0.761} & -0.036 \\
   ${\rm log}\,c_\mathrm{vir}$ & {\bf 0.630} & 0.059 & 0.247 & 0.049 \\
   ${\rm log}\,\lambda$ & {\bf -0.499} & -0.054 & 0.123 & 0.226 \\
   ${\rm log}\,q$ & -0.126 & {\bf -0.585} & 0.115 & {\bf -0.667} \\
   ${\rm log}\,\Phi_L$ & -0.007 & {\bf -0.595} & {\bf -0.495} & 0.159 \\
   ${\rm log}\,x_\mathrm{off}$ & {\bf 0.509} & 0.205 & -0.149 & 0.187 \\
   ${\rm log}\,\rho_\mathrm{rms}$ & 0.128 & {\bf -0.494} & 0.256 & {\bf 0.664} \\
   \hline
   eigenvalues & 1.51 & 1.10 & 1.02 & 0.97 \\
   \% of variance & 21.6\% & 15.7\% & 14.6\% & 13.8\% \\
  \end{tabular}
 \end{center}
 \caption{`Good' (relaxed) haloes with ${\rm log}~(M/h^{-1}M_\odot)\sim13.3$ and more than 1000 particles, 180~Mpc box. $N$=383.}
 \label{good180delta}
 \begin{center}
 \begin{tabular}[h!]{ l | c c c c }
   property & PC 1 & PC 2 & PC 3 & PC 4 \\
   \hline
   ${\rm log}\,\Delta_8$ & -0.138 & -0.164 & {\bf 0.819} & 0.418 \\
   ${\rm log}\,c_\mathrm{vir}$ & {\bf 0.629} & 0.051 & 0.112 & -0.061 \\
   ${\rm log}\,\lambda$ & {\bf -0.521} & -0.050 & 0.095 & 0.138 \\
   ${\rm log}\,q$ & -0.050 & {\bf 0.603} & 0.249 & -0.149 \\
   ${\rm log}\,\Phi_L$ & 0.093 & {\bf 0.405} & -0.347 & {\bf 0.840} \\
   ${\rm log}\,x_\mathrm{off}$ & {\bf -0.544} & 0.256 & -0.206 & -0.183 \\
   ${\rm log}\,\rho_\mathrm{rms}$ & 0.084 & {\bf 0.613} & 0.288 & -0.203 \\
   \hline
   eigenvalues & 1.50 & 1.11 & 1.02 & 0.96 \\
   \% of variance & 21.5\% & 15.9\% & 14.6\% & 13.8\% \\
  \end{tabular}
 \end{center}
 \caption{`Good' (relaxed) haloes with ${\rm log}~(M/h^{-1}M_\odot)\sim13.6$ and more than 1000 particles, 300~Mpc box. $N$=944.}
 \label{good300delta}
\end{table}

We find that the local overdensity is usually significant on the third principle component, 
where it is sometimes paired with the angle $\Phi_L$.  
This is consistent with Altay et al.\ (2006), who found that haloes are aligned 
with their large-scale structures (if they are located in filaments, for example), 
while halo shapes are not dependent on membership of a filament. 
It is also consistent with Wang et al.\ (2011), who showed that a halo's spin is correlated 
with the local overdensity and the strength of the tidal field. 
In any case, in all of our simulations, 
the density is rarely significant on the first two PCs.
Therefore, we argue that, at fixed mass, a halo's environment is relatively unimportant, 
because it does not significantly determine the halo's properties.  At fixed mass, 
a halo's angular momentum and degree of relaxedness are much more important.

\subsection{Correlation with $\Phi_L$}
\label{sec:phil}

%
%



\begin{figure}
\includegraphics[width=0.5\textwidth]{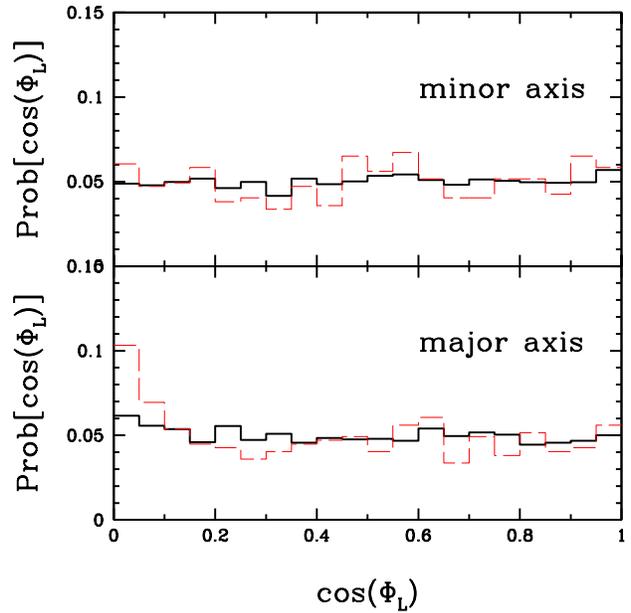}
\caption{
         Probability distribution for the angle $\Phi_L$ between the angular 
         momentum vector $L$ and the minor (upper panel) and major (lower panel) 
         axis of the dark matter distribution. The solid black line is for 
         relaxed haloes in the 90~Mpc box, while the red line is for the spin 
         haloes ($\lambda > 0.07$, $\approx 12\%$ of the total).
        }
\label{fig:phildist}
\end{figure}

As previously stated the angle $\Phi_L$ can be thought of as a proxy for the 
position of a possible stellar disk within the dark matter halo, assuming an 
alignment between the angular momentum of the dark matter particle and the 
stellar component (e.g., Hahn et al.\ 2010). Figure \ref{fig:phildist} shows 
the probability distribution of the cosine of $\Phi_L$ in the high resolution 
90~Mpc box.  In the upper panel the angle $\Phi_L$ is defined as the angle 
between the minor axis and the angular momentum $L$, while in the lower panel 
the major axis is used.  For relaxed haloes the probability distribution is 
remarkably flat in both cases, and there is no sign of a possible alignment 
between either the major or minor axis and $L$.  If we restrict our analysis 
to high spin haloes ($\lambda > 0.07$, red dashed line, $\approx 12\%$ of the 
total) a weak maximum at $\cos (\Phi_L)=0$ arises for the angle with the 
major axis, while there is still no evidence for a correlation between the 
orientation of the minor axis and the halo angular momentum. 

%
%

This result is consistent with Zentner et al.\ (2005b) and Kuhlen et al.\ 
(2007), who showed that subhaloes are distributed anisotropically and are 
preferentially aligned with the major axis of the triaxial halo.  Studies of 
satellite galaxies in groups have reached similar conclusions (Yang et al.\ 2006; 
Faltenbacher et al.\ 2007; Kang et al.\ 2007; Wang et al.\ 2008). 
Nonetheless, other studies have detected significant correlations between the 
spin axis of haloes and their \textit{minor} axis, stronger than correlations 
with the major axis (Bailin 
\& Steinmetz 2005; Bett et al.\ 2007).  It is possible that the disagreement 
between our result (Fig.~\ref{fig:phildist}) and these studies somehow owes to 
the differences between SO and friends-of-friends (FOF) halo-finding algorithms.  
As noted by a number of authors, there are different disadvantages to these 
two types of halo-finders (e.g., Bett et al.\ 2007; Tinker et al.\ 2008b; Knebe 
et al.\ 2011).  
In particular, SO halo-finders tend to impose a more spherical geometry on the 
resulting systems.  On the other hand, FOF halo-finders often identify objects linked with 
neighboring objects via tenuous bridges of particles, resulting in anomolously large velocity dispersions, masses, and spin parameters of $\approx 6\%$ of objects (Bett et al.\ 2007). 
In addition, FOF halo-finders define a halo center as the center-of-mass, while most 
other algorithms (including ours) define the center as the center of the density profile 
(i.e., the most bound particle) (Knebe et al.\ 2011).  These differences with FOF 
halo-finders may explain the disagreement about correlations between the spin axis 
and major or minor axes of haloes. 

Returning to our PCA results, 
in the different simulation boxes, rarely does a single halo parameter 
dominate a particular principle component. 
Nonetheless, $\Phi_L$ usually dominates the third or fourth PC, 
although sometimes the shape parameter is also significant on the same PC. 
In other words, one might interpret this as evidence that $\Phi_L$ is weakly correlated 
with spin and/or shape, but not related to other halo properties.
$\Phi_L$, which is the  angle between the major axis of the halo and its angular momentum vector,
is expected to be important for the formation of disk galaxies, because 
the angular momenta of the stellar disk and the dark matter particles are generally, 
though not always, expected to be aligned (van den Bosch et al.\ 2002; Kazantzidis et al.\ 2004; 
Zentner et al.\ 2005b; Sharma \& Steinmetz 2005; Bett et al.\ 2010; Hahn et al.\ 2010).

\begin{figure}
\includegraphics[width=0.5\textwidth]{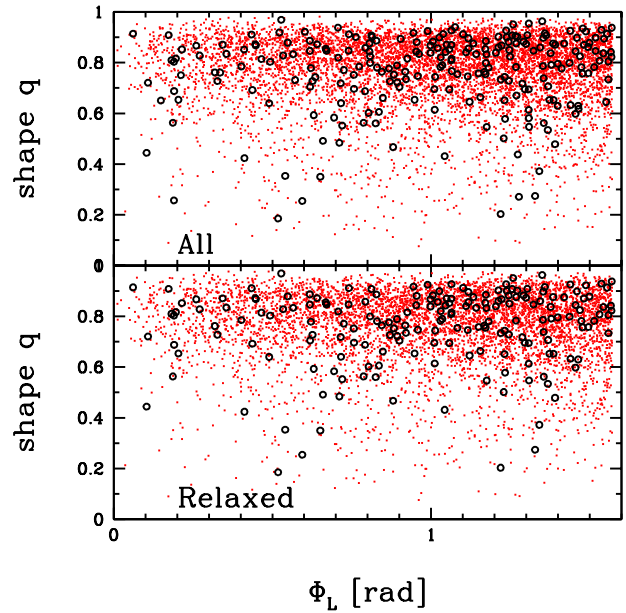}
\caption{Relation between the shape ($q$) and the angle $\Phi_L$.
Black circle and red dots represent haloes from the W5-30.1 and the W5-90.2 simulations,  respectively.
The two panel shows results for All (upper) and Relaxed haloes. In both cases there is no correlation between the two properties.}
\label{fig:shape3}
\end{figure}

\begin{figure}
\includegraphics[width=0.5\textwidth]{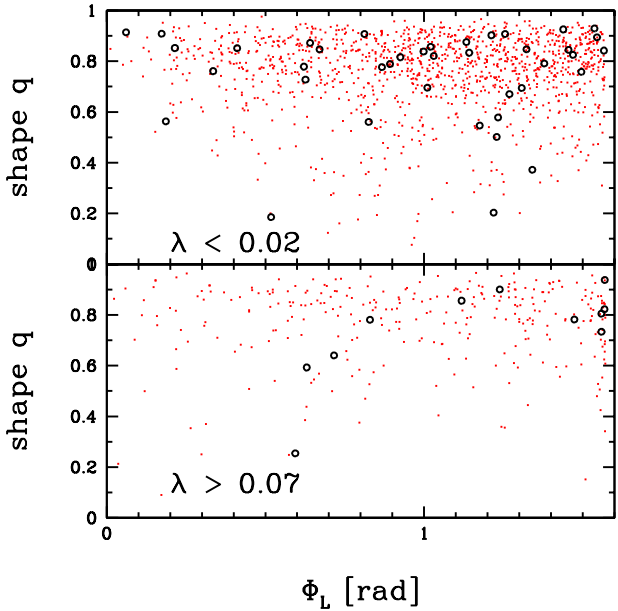}
\caption{
Same as Figure \ref{fig:shape3} but of haloes with different spins.
The upper panel shows results for relaxed haloes with little angular 
momentum (defined as haloes with $\lambda<0.02$) while the lower one 
is for relaxed haloes with a lot of angular momentum ($\lambda>0.07$).
Again, there is no correlation between spin ($\lambda$) and the angle $\Phi_L$.}
\label{fig:spin2}
\end{figure}

In contrast to this, we show in Figure~\ref{fig:shape3} 
that $\Phi_L$ is uncorrelated with the shape parameter $q$.
This result is independent of the value of the angular momentum
(quantified by the spin parameter) as shown in Figure~\ref{fig:spin2}. 
Therefore, as a consequence of our results, 
we argue that it is not possible to reliably determine the angle of 
the angular momentum vector from a halo's shape. 
This has consequences for the formation of disk galaxies, which we discuss in the 
final section of the paper. 

\section{Dependence on Halo Relaxedness}
\label{sec:relax}


From the results of the PCAs in Sections~\ref{sec:Mmin} and \ref{sec:fixedM}, 
we argued that halo `relaxedness' is as important as halo mass in determining halo 
properties, because mass and relaxedness account for the majority of the variance in the PCAs.  
This conclusion about the importance of halo relaxedness is consistent 
with other studies, such as Shaw et al.\ (2006), who find that the dynamical state of 
haloes, quantified by the virial ratio, is correlated with halo parameters: mass, 
concentration, spin as well as the substructure fraction.  In addition, Neto et al.\ (2007) 
define an `equilibrium state' in terms of the substructure fraction, center of mass 
displacement (like our $x_\mathrm{off}$), and virial ratio, and find that a halo's 
equilibrium state is correlated with its mass, concentration, and spin.

In order to investigate this issue further, in this section we quantify and 
compare correlations between halo properties and mass, and between them and the 
relaxedness parameters.

\begin{figure*}
\includegraphics[width=0.497\textwidth]{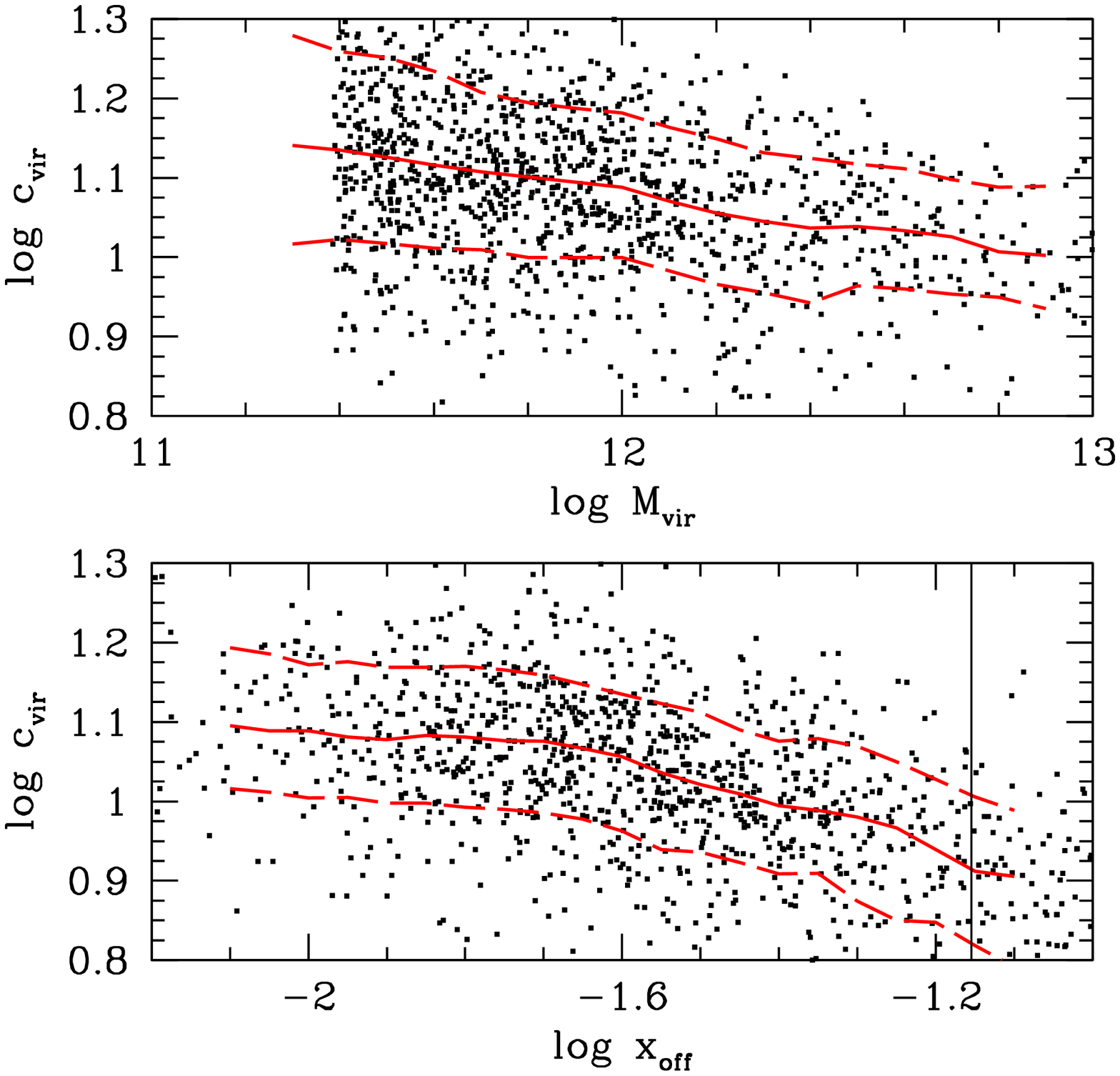}
\includegraphics[width=0.497\textwidth]{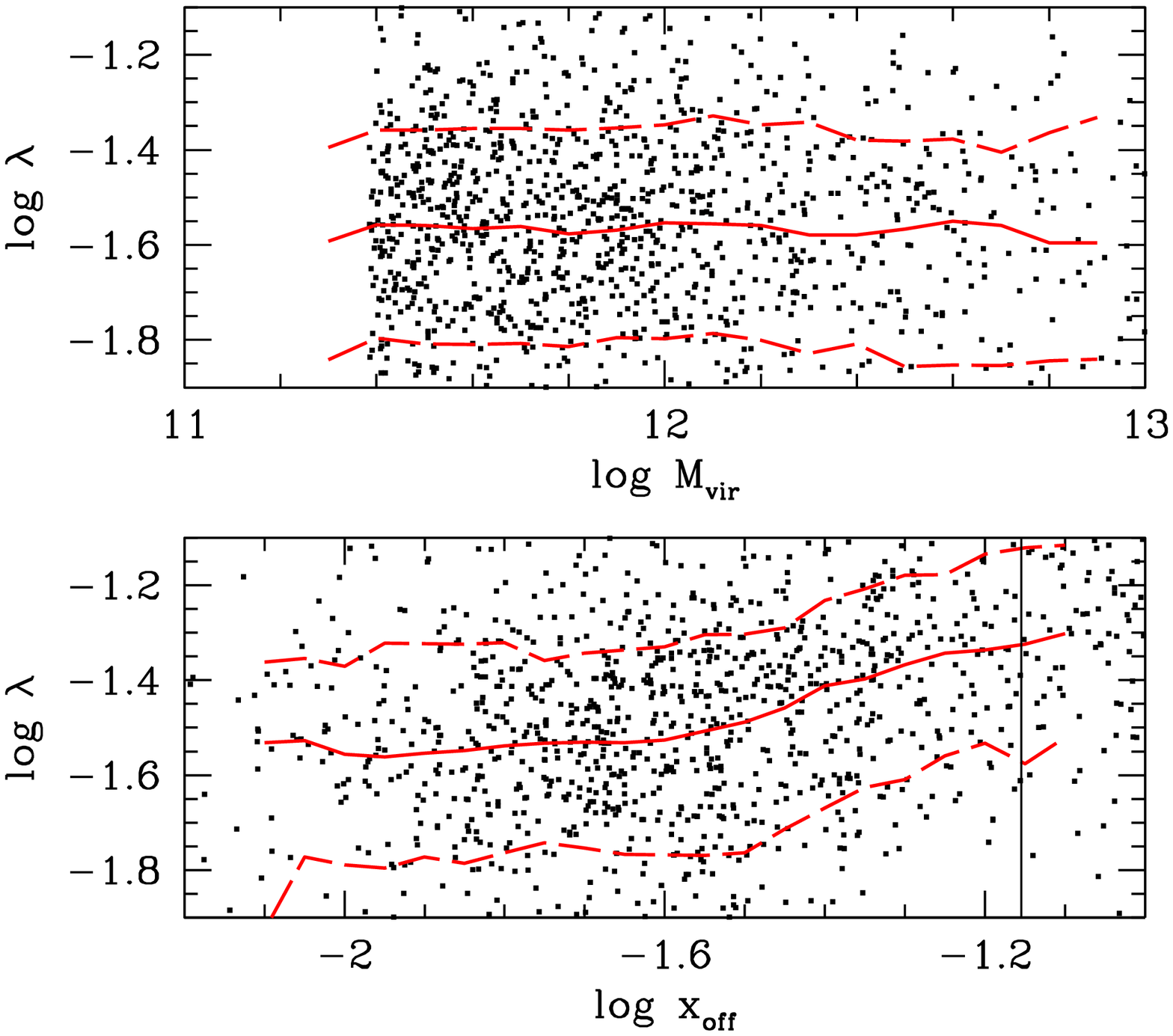}
\caption{Left: $M_\mathrm{vir}$ vs $c_\mathrm{vir}$ for very relaxed haloes 
         ($x_\mathrm{off}<0.018$, $\rho_\mathrm{rms}<0.14$) 
         in the upper panel, and $x_\mathrm{off}$ vs $c_\mathrm{vir}$ at fixed mass 
         ($11.85\leq \mathrm{log}~(M/h^{-1}M_\odot)<12.15$, as in Section~\ref{sec:fixedM}) in the 
         lower panel, for haloes with more than 1000 particles in the 90~Mpc box. 
         Right: same, but with the spin parameter, $\lambda$.  
         Points indicate individual haloes, solid lines show the running 
         medians, and dashed lines show the 16 and 84 percentiles. 
         The vertical lines in the lower panels indicate our cut for relaxed haloes, at $x_\mathrm{off}<0.07$.
        }
\label{Mvsrelaxplots}
\end{figure*}

In particular, we first perform Spearman rank tests\footnote{The Spearman rank correlation coefficient may have a value between $-1$ and 1.  A positive (negative) value indicates an (anti)correlation, and a value of 0 indicates no correlation.} on 
some correlations between these parameters. 
We begin with correlations with halo relaxedness parameters at fixed mass ($11.85\leq \mathrm{log}~(M/h^{-1}M_\odot)<12.15$), using haloes with more than 1000 particles in the 90~Mpc box.  This is one of the catalogues used in Section~\ref{sec:fixedM} (see PCA result in Table~\ref{good90M}). The correlations between concentration and $x_\mathrm{off}$, and between spin and $x_\mathrm{off}$ are statistically significant, with Spearman ranks $r_s=-0.65$ and $0.42$, respectively, and remain significant if only `good' (i.e., relaxed) haloes are selected.  The correlation with shape is weaker, while that with $\Phi_L$ is not significant at all. 
Of the correlations with $\rho_\mathrm{rms}$, only that with concentration is significant, with $r_s=-0.45$.  Therefore, the dependence on relaxedness at fixed mass is predominantly due to $x_\mathrm{off}$.

In order to perform a fair comparison with the halo mass dependence, we test the strength of correlations with mass at fixed relaxedness: we use the same simulation box and select haloes within a narrow range of relaxedness parameters ($x_\mathrm{off}<0.018$ and $\rho_\mathrm{rms}<0.14$), such that there are a similar number of haloes ($\sim1300$) as the previous test at fixed mass.  Only the resulting mass-concentration correlation is still significant, with $r_s=-0.38$.  These results imply that the expected weak correlations between mass and the spin parameter, and between mass and shape (e.g., Macci\`{o} et al.\ 2007; Bett et al.\ 2007), are no longer significant once one accounts for the dependence on halo relaxedness.

In Figure~\ref{Mvsrelaxplots}, we show some of these correlations that we tested with the Spearman rank coefficients.  In the left panels of Figure~\ref{Mvsrelaxplots}, we show the correlation between concentration and $x_\mathrm{off}$ at fixed mass, and compare it to the mass-concentration relation at fixed relaxedness parameters (similar to Fig.~2 of Macci\`{o} et al.\ (2008)).  Halo mass and the center of mass offset appear to be independently correlated with concentration (i.e., one is not due to the other).  More massive haloes and haloes with larger offsets tend to be less concentrated.

We show the analogous plots with the spin parameter $\lambda$ in the right panels of Figure~\ref{Mvsrelaxplots}. One can see no mass dependence at fixed relaxedness, although at fixed mass, haloes with larger offsets $x_\mathrm{off}$ tend to have larger spin.  Although the spin parameter is not well-defined for unrelaxed haloes, we see that the correlation occurs for haloes defined as relaxed (with $\mathrm{log}~x_\mathrm{off}<-1.15$, or the stricter criterion in Macci\`{o} et al.\ (2007), $\mathrm{log}~x_\mathrm{off}<-1.40$). 

These tests show that halo concentration and spin are as tightly correlated with the relaxedness parameters (especially $x_\mathrm{off}$) as with halo mass.  This confirms our conclusion that the relaxed state of a halo is as important as its mass, in determining its other properties.

\section{Discussion and conclusions}
\label{sec:concl}

%

We used numerical simulations of dark matter haloes to study a variety of halo 
properties: mass, concentration, spin, shape, the angle between the major axis and 
the angular momentum vector ($\Phi_L$), the distance between the most bound particle 
and the centre of mass ($x_\mathrm{off}$), the r.m.s. of the NFW fit to the 
density profile ($\rho_\mathrm{rms}$), and the local overdensity. 
We analyzed all of these properties together, by employing principal component analysis (PCA).

We summarize our main results as follows: 

\begin{itemize}

\item
There is no single dominant PC or pair of PCs.  Unlike the spectra of galaxies, 
for dark matter haloes, there is no parameter or combination of parameters that 
strongly determine the other parameters..

\item
We find that whether a halo is relaxed or not 
is at least as important as the halo's mass and concentration. 
Even for relatively relaxed haloes, the degree of relaxedness, quantified by 
$x_\mathrm{off}$ and $\rho_\mathrm{rms}$, is still as important as mass and 
concentration, and these four properties tend to dominate the first two principal components.

\item 
$\Phi_L$ is not significantly correlated with any other halo properties. 
It is therefore not possible to reliably estimate the direction of the 
angular momentum vector of a halo from its shape.

\item
A halo's `environment' (quantified by the 8 Mpc overdensity) usually dominates 
the third principle component, where it is sometimes combined with the angle 
$\Phi_L$, but it is less significant than other halo properties on the first two 
principle components.  Therefore, at fixed halo mass, the environment is 
relatively unimportant in determining the other halo properties.

\end{itemize}

Our results have many important implications, especially for halo occupation models 
and semi-analytic models.
Most such models explicitly assume that halo mass (or the combination of mass and concentration) 
largely determines other important halo properties, and to some extent determines the properties 
of the galaxies hosted by a halo; 
however, our results show that this assumption is often false: halo mass is less dominant 
than one might expect.  \textit{The degree of relaxedness of a halo is at least as important as its mass.}  
Therefore, the relaxedness of a halo could affect the formation of galaxies hosted by it, 
independent of halo mass.  In an unrelaxed halo, for example, galaxies may experience stronger 
tidal forces or their supply of gas to form stars could be disrupted. 
For semi-analytic models, one solution may be to quantify the relaxedness of the dark matter 
haloes and to use them in addition to the halo merger histories.  For halo occupation models, it 
is possible that the unrelaxedness of some haloes may affect or blur the relation between halo 
mass and central galaxy luminosity or stellar mass.  
In addition, it is possible that the halo mass function and halo mass-concentration 
relation assumed in the models could be different for relaxed and unrelaxed haloes.
It is not yet clear how strongly halo 
relaxedness may affect results from halo occupation and semi-analytic models, and 
these issues deserve further study.

A related issue is that of halo `assembly bias', in which the formation time of haloes (or 
other properties related to halo assembly) is correlated with the environment at fixed mass 
(e.g., Sheth \& Tormen 2004; Croton, Gao \& White 2007; Wetzel et al.\ 2007; Faltenbacher \& White 2010; Fakhouri \& Ma 2010).  
The relaxedness of haloes, which we 
quantify with $\rhorms$ and $\xoff$, is clearly related to their assembly: a halo that has 
recently formed or recently experienced a merger is more likely to be unrelaxed.  Although 
our density parameter, $\Delta_8$ is subdominant in our PCAs, it is nonetheless possible that 
the relaxedness of haloes is related to their larger scale environment.  Hence, the correlations 
with halo relaxedness may be a manifestation of halo assembly bias.  Nonetheless, assembly bias 
may have a relatively weak effect for halo models of galaxy clustering (e.g., Blanton \& 
Berlind 2007; Tinker et al.\ 2008a; Skibba \& Sheth 2009).

Our result for the angle $\Phi_L$ also has interesting implications.  The fact that 
$\Phi_L$ is not significantly correlated with other halo properties may present a 
difficulty for some models.  While the angular momentum of the stellar disk of a 
galaxy is thought to be aligned with the dark matter particles of the host halo 
(such that the pole of the disk is collinear with the angular momentum vector of the halo), 
\textit{our result implies that it is actually quite difficult, if not impossible, 
to accurately determine the orientation of the disk simply by looking at the halo's shape.}
This is consistent with hydrodynamical simulations that include baryonic cooling, which result in disks with spin axes that are very poorly aligned with the halo (Bailin et al.\ 2005). 
It suggests that the direction of the angular momentum of the baryons is not well conserved 
throughout the disk formation process (van den Bosch et al.\ 2002; Yang et al.\ 2006). 
Heller et al.\ (2007) also argue that the shape and morphology of a disk depend on the 
relative angle with the major axis of the halo, but this dependence too is not well conserved. 
Therefore, although the formation of disk galaxies and their host halos are certainly 
related, after they have evolved, it is not clear that one could reliably determine to what 
extent the disk remains aligned with the halo.

Finally, at the time of publication, Jeeson-Daniel et al.\ (2011) posted a preprint 
of a related study, involving PCAs of various halo properties.  Some of the results in 
Jeeson-Daniel et al.\ are consistent with ours, such as that halo mass, concentration, 
substructure, and relaxedness are among a set of parameters that together account for a 
large fraction of the variance.  Nonetheless, their relaxedness parameter, which is 
similar to our $\xoff$, accounts for less variance in their PCAs than in ours.

\section*{Acknowledgments} 

We thank Carlo Giocoli, Ravi Sheth, Kester Smith, Vivi Tsalmantza, and Frank van den Bosch 
for valuable discussions that helped to improve the quality of this paper.
We also thank the anonymous referee for comments that helped to clarify and strengthen the paper's arguments. 
Numerical simulations were performed on the PIA and on PanStarrs2 clusters of the
Max-Planck-Institut f\"ur Astronomie at the Rechenzentrum in Garching.


\appendix

\renewcommand{\thefigure}{\Alph{appfig}\arabic{figure}}
\setcounter{appfig}{1}

\section{Supplementary PCAs and Tests}\label{markdistapp}

%
%
%
%
%

In this appendix, we include additional principle component analyses that 
can be compared to the results in Section~\ref{sec:res}.

Firstly, we perform PCAs for halo mass threshold ($\mathrm{log}~(M/h^{-1}M_\odot)>12$) for 
the 40 and 300~Mpc simulation boxes, analogous to the results shown in 
Tables~\ref{good90} and \ref{good180}.
The results of these PCAs are shown in Tables~\ref{good40} and \ref{good300}. 
The PCAs of the 40 and 90~Mpc boxes are similar, although for the 40~Mpc 
box, $\lambda$ is now very significant on the second PC. 
The PCAs of the 180 and 300~Mpc boxes are similar as well, although for the 300 Mpc 
box, $c_\mathrm{vir}$ is more significant on the first PC than on the second one. 
In any case, even when one compares the PCAs of the 40 and 300~Mpc boxes, 
which have very different resolutions, the results are quite similar.  
We conclude that our results are not very sensitive to the resolution of the simulations.
\begin{table}
 \begin{center}
 \begin{tabular}[h!]{ l | c c c c }
   property & PC 1 & PC 2 & PC 3 & PC 4 \\
   \hline
   ${\rm log}\,M_{\rm vir}$ & {\bf -0.618} & 0.078 & -0.002 & 0.015 \\
   ${\rm log}\,c_\mathrm{vir}$ & {\bf 0.408} & {\bf 0.499} & -0.015 & 0.078 \\
   ${\rm log}\,\lambda$ & -0.068 & {\bf -0.645} & -0.046 & -0.070 \\
   ${\rm log}\,q$ & 0.093 & -0.267 & 0.458 & {\bf 0.820} \\
   ${\rm log}\,\Phi_L$ & 0.002 & -0.053 & {\bf -0.881} & 0.467 \\
   ${\rm log}\,x_\mathrm{off}$ & 0.292 & {\bf -0.503} & -0.106 & -0.313 \\
   ${\rm log}\,\rho_\mathrm{rms}$ & {\bf 0.595} & -0.046 & -0.014 & -0.022 \\
   \hline
   eigenvalues & 2.14 & 1.26 & 1.01 & 0.97 \\
   \% of variance & 30.6\% & 18.0\% & 14.4\% & 13.9\% \\
  \end{tabular}
 \end{center}
 \caption{Relaxed haloes with ${\rm log}~(M/h^{-1}M_\odot)\geq12$, 40~Mpc box. $N$=1822.}
 \label{good40}
 \begin{center}
 \begin{tabular}[h!]{ l | c c c c }
   property & PC 1 & PC 2 & PC 3 & PC 4 \\
   \hline
   ${\rm log}\,M_{\rm vir}$ & {\bf -0.460} & 0.472 & -0.124 & -0.003 \\
   ${\rm log}\,c_\mathrm{vir}$ & {\bf 0.609} & 0.185 & 0.045 & -0.009 \\
   ${\rm log}\,\lambda$ & -0.338 & {\bf -0.465} & 0.116 & 0.044 \\
   ${\rm log}\,q$ & 0.075 & -0.367 & -0.360 & {\bf -0.818} \\
   ${\rm log}\,\Phi_L$ & 0.112 & -0.161 & {\bf -0.862} & 0.463 \\
   ${\rm log}\,x_\mathrm{off}$ & -0.312 & {\bf -0.483} & 0.136 & 0.240 \\
   ${\rm log}\,\rho_\mathrm{rms}$ & {\bf 0.434} & -0.364 & 0.279 & 0.239 \\
   \hline
   eigenvalues & 1.62 & 1.33 & 0.99 & 0.96 \\
   \% of variance & 23.2\% & 19.0\% & 14.2\% & 13.7\% \\
  \end{tabular}
 \end{center}
 \caption{Relaxed haloes with ${\rm log}~(M/h^{-1}M_\odot)\geq12$ and more than 1000 particles, 300~Mpc box. $N$=1905.}
 \label{good300}
\end{table}



Next, we perform more PCAs at fixed mass, using 
smaller-scale overdensities $\Delta_2$ and $\Delta_4$, in 2 and 4 Mpc apertures, 
analogous to the results in Table~\ref{good180delta}.
The results of these PCAs are shown in Tables~\ref{good180delta2} and \ref{good180delta4}. 
When using $\Delta_4$, the first three PCs are essentially the same as when $\Delta_8$ was used, 
and $q$ still dominates the fourth PC.  When $\Delta_2$ is used, the second and third 
PCs are different, although they are dominated by the same parameters.  In any case, 
regardless of the scale of the overdensity, we conclude that halo environment is less 
prominent in our PCAs, and therefore less important, than the other halo parameters.
\begin{table}
 \begin{center}
 \begin{tabular}[h!]{ l | c c c c }
   property & PC 1 & PC 2 & PC 3 & PC 4 \\
   \hline
   ${\rm log}\,\Delta_2$ & 0.282 & {\bf 0.406} & {\bf 0.559} & 0.031 \\
   ${\rm log}\,c_\mathrm{vir}$ & {\bf -0.621} & 0.221 & 0.150 & -0.023 \\
   ${\rm log}\,\lambda$ & {\bf 0.511} & 0.035 & 0.181 & -0.290 \\
   ${\rm log}\,q$ & 0.098 & {\bf -0.542} & 0.201 & {\bf 0.755} \\
   ${\rm log}\,\Phi_L$ & -0.023 & {\bf -0.688} & 0.084 & {\bf -0.580} \\
   ${\rm log}\,x_\mathrm{off}$ & {\bf 0.505} & 0.061 & -0.201 & 0.062 \\
   ${\rm log}\,\rho_\mathrm{rms}$ & -0.098 & -0.118 & {\bf 0.738} & -0.070 \\
   \hline
   eigenvalues & 1.52 & 1.11 & 1.08 & 0.97 \\
   \% of variance & 21.8\% & 15.8\% & 15.4\% & 13.8\% \\
  \end{tabular}
 \end{center}
 \caption{Relaxed haloes with ${\rm log}~(M/h^{-1}M_\odot)\sim13.3$ and more than 1000 particles, 180~Mpc box. $N$=383.}
 \label{good180delta2}
 \begin{center}
 \begin{tabular}[h!]{ l | c c c c }
   property & PC 1 & PC 2 & PC 3 & PC 4 \\
   \hline
   ${\rm log}\,\Delta_4$ & 0.233 & -0.181 & {\bf 0.751} & 0.049 \\
   ${\rm log}\,c_\mathrm{vir}$ & {\bf -0.626} & 0.024 & 0.303 & -0.024 \\
   ${\rm log}\,\lambda$ & {\bf 0.502} & -0.066 & 0.157 & -0.271 \\
   ${\rm log}\,q$ & 0.124 & {\bf -0.569} & -0.110 & {\bf 0.772} \\
   ${\rm log}\,\Phi_L$ & 0.131 & {\bf -0.572} & {\bf -0.442} & {\bf -0.470} \\
   ${\rm log}\,x_\mathrm{off}$ & {\bf 0.521} & 0.200 & -0.006 & -0.037 \\
   ${\rm log}\,\rho_\mathrm{rms}$ & -0.123 & {\bf -0.521} & 0.336 & -0.325 \\
   \hline
   eigenvalues & 1.51 & 1.10 & 1.03 & 0.97 \\
   \% of variance & 21.6\% & 15.7\% & 14.7\% & 13.8\% \\
  \end{tabular}
 \end{center}
 \caption{`Good' (relaxed) haloes with ${\rm log}~(M/h^{-1}M_\odot)\sim13.3$ and more than 1000 particles, 180~Mpc box. $N$=383.}
 \label{good180delta4}
\end{table}

Finally, while we have attempted to minimize the effects of correlated errors, we 
have not heretofore accounted for parameters with particularly large errors.  
One way to test the effects of such errors is to add error (with a Gaussian 
distribution) to a particular parameter.  Such tests yield PCA results similar 
to those shown in Section~\ref{sec:res}, although we find that the first few principal 
components are usually more robust than the later PCs.
An example is shown in Table~\ref{good180sigma}, in which we have significantly increased the error 
of the spin parameter, by 1-$\sigma$.   Comparing to Table~\ref{good180delta},
the numbers are slightly different, but the dominant parameters on each PC remain the same. 
The only exception is $\lambda$ itself, which is no longer significant on PC~1, but is 
instead slightly significant on PC~2 and more significant on PC~4.
Similar tests with other parameters have yielded smaller changes than shown here: 
for example, adding error to the concentration does not affect any of the dominant 
parameters on the PCs, and only slightly decreases $c_\mathrm{vir}$'s significance on PC~1. 
We have also tested the effects of asymmetric errors, but these significantly affect the PCA results only if they are larger and much more asymmetric than the expected errors in these simulations. 
We conclude that our PCA results are robust to the parameter errors. 
\begin{table}
 \begin{center}
 \begin{tabular}[h!]{ l | c c c c }
   property & PC 1 & PC 2 & PC 3 & PC 4 \\
   \hline
   ${\rm log}\,\Delta_8$ & 0.249 & 0.100 & {\bf 0.762} & 0.210 \\
   ${\rm log}\,c_\mathrm{vir}$ & {\bf -0.663} & 0.022 & 0.195 & 0.176 \\
   ${\rm log}\,\lambda$ & 0.282 & 0.367 & -0.020 & {\bf 0.639} \\
   ${\rm log}\,q$ & 0.186 & {\bf 0.428} & 0.251 & {\bf -0.713} \\
   ${\rm log}\,\Phi_L$ & 0.034 & {\bf 0.623} & {\bf -0.504} & 0.034 \\
   ${\rm log}\,x_\mathrm{off}$ & {\bf 0.602} & -0.245 & -0.084 & 0.043 \\
   ${\rm log}\,\rho_\mathrm{rms}$ & -0.143 & {\bf 0.472} & 0.240 & 0.073 \\
   \hline
   eigenvalues & 1.38 & 1.12 & 1.02 & 1.00 \\
   \% of variance & 19.7\% & 16.0\% & 14.5\% & 14.3\% \\
  \end{tabular}
 \end{center}
 \caption{`Good' (relaxed) haloes with ${\rm log}~(M/h^{-1}M_\odot)\sim13.3$ and more than 1000 particles, 180~Mpc box. $N$=383.  Like Table~\ref{good180delta}, but with 1-$\sigma$ error added to ${\rm log}\,\lambda$.}
 \label{good180sigma}
\end{table}

\label{lastpage}


\begin{thebibliography}{99}





\bibitem[{Abbas \& Sheth (2007)}]{as07}
 Abbas U., Sheth R. K., 2007, MNRAS, 378, 641

\bibitem[{Agertz et al. (2010)}]{agertz}
 Agertz O., Teyssier R., Moore B., 2011, MNRAS, 410, 1391

\bibitem[{Allgood et al. (2006)}]{allgood06}
 Allgood B., Flores R. A., Primack J. R., Kravtsov A. V., Wechsler R. H., Faltenbacher A., Bullock J. S., 2006, MNRAS, 367, 1781

\bibitem[{Altay et al. (2006)}]{altay}
 Altay G., Colberg J., Croft R. A. C., 2006, MNRAS, 370, 1422

\bibitem[{Avila-Reese et al. (2005)}]{ar05}
 Avila-Reese V., Col\'{i}n P., Gottl\"{o}ber S., Firmani C., Maulbetsch C., 2005, ApJ, 634, 51

\bibitem[{Bailer-Jones et al. (1998)}]{bj}
 Bailer-Jones C. A. L., Irwin M., von Hippel T., 1998, MNRAS, 298, 361

\bibitem[{Bailin et al. (2005)}]{bailin05}
 Bailin J., et al., 2005, ApJ, 627, L17

\bibitem[{Bailin \& Steinmetz (2005)}]{bailin}
 Bailin J., Steinmetz M., 2005, ApJ, 627, 647

\bibitem[{Bett et al. (2007)}]{bett07}
 Bett P., Eke V., Frenk C. S., Jenkins A., Helly J., Navarro J., 2007, MNRAS, 376, 215

\bibitem[{Bett et al. (2010)}]{bett10}
 Bett P., Eke V., Frenk C. S., Jenkins A., Okamoto T., 2010, MNRAS, 404, 1137


\bibitem[{Bertschinger (2001)}]{bert}
 Bertschinger E., 2001, ApJS, 137, 1

\bibitem[{Blanton \& Berlind (2007)}]{bb07}
 Blanton M. R., Berlind A. A., 2007, ApJ, 664, 791

\bibitem[{Bonoli \& Pen (2009)}]{bonoli}
 Bonoli S., Pen U. L., 2009, MNRAS, 396, 1610 

\bibitem[{Boroson \& Lauer (2010)}]{bl10}
 Boroson T. A., Lauer T. R., 2010, AJ, 140, 390 

\bibitem[{Bower et al. (2006)}]{bower05} 
 Bower R. G., Benson A. J., Malbon R., Helly J. C., Frenk C.  S., Baugh C.  M., Cole S., Lacey C. G., 2006, MNRAS, 370, 645

\bibitem[{Bullock et al. (2001a)}]{b01}
 Bullock J. S., Kolatt T. S., Sigad Y., Somerville R. S., Kravtsov A. V., Klypin A. A., Primack J. R., Dekel A., 2001, MNRAS, 321, 559

\bibitem[{Cattaneo et al. (2007)}]{cattaneo}
 Cattaneo A., et al., 2007, MNRAS, 377, 63

\bibitem[{Chen et al. (2009)}]{chen}
 Chen Y.-M., Wild V., Kauffmann G., Blaizot J., Davis M., Noeske K., Wang J.-M., Willmer C., 2009, MNRAS, 393, 406

\bibitem[{Colberg et al. (1999)}]{colberg}
 Colberg J. M., White S. D. M., Jenkins A., Pearce F. R., 1999, MNRAS, 308, 593

\bibitem[{Connolly et al. (1995)}]{connolly}
 Connolly A. J., Szalay A. S., Bershady M. A., Kinney A. L., Calzetti D., 1995, AJ, 110, 1071

\bibitem[{Connolly \& Szalay (1999)}]{cs99}
 Connolly A. J., Szalay A. S., 1999, AJ, 117, 2052

\bibitem[{Conselice (2006)}]{conselice}
 Conselice C. J., 2006, MNRAS, 373, 1389

\bibitem[{Croton et al. (2006)}]{croton}
 Croton D. J., Gao L., White S. D. M., 2007, MNRAS, 374, 1303

\bibitem[{Dalal et al. (2008)}]{dalal}
 Dalal N., White M., Bond J. R., Shirokov A., 2008, ApJ, 687, 12


\bibitem[{De Lucia \& Blaizot (2007)}]{delucia}
 De Lucia G., Blaizot J., 2007, MNRAS, 375, 2


\bibitem[{Eisenstein et al. (2003)}]{eis03}
 Eisenstein D. J., et al., 2003, ApJ, 585, 694

\bibitem[{Efstathiou \& Fall (1984)}]{ef84}
 Efstathiou G., Fall M. S., 1984, MNRAS, 206, 453

\bibitem[{Faber (1973)}]{faber}
 Faber S. M., 1973, ApJ, 179, 731

\bibitem[{Fakhouri \& Ma (2010)}]{fm10}
 Fakhouri O., Ma C.-P., 2010, MNRAS, 401, 2245

\bibitem[{Faltenbacher et al. (2007)}]{falt07}
 Faltenbacher A., Li C., Mao S., van den Bosch F. C., Yang X., Jing Y. P., Pasquali A., Mo H. J., 2007, ApJ, 662, L71

\bibitem[{Faltenbacher \& White (2010)}]{fw10}
 Faltenbacher A., White S. D. M., 2010, ApJ, 708, 469

\bibitem[{Ferreras et al. (2006)}]{ferreras}
 Ferreras I., Pasquali A., de Carvalho R. R., de la Rosa I. G., Lahav O., 2006, MNRAS, 370, 828

\bibitem[{Gao et al. (2005)}]{gao}
 Gao L., Springel V., White S. D. M., 2005, MNRAS, 363, L66

\bibitem[{Giocoli et al. (2009)}]{giocoli}
 Giocoli C., Tormen G., Sheth R. K., van den Bosch F. C., 2010, MNRAS, 404, 502 

\bibitem[{Governato et al. (2010)}]{governato}
 Governato F., et al., 2010, Nature, 463, L203

\bibitem[{Hahn et al. (2010)}]{hahn}
 Hahn O., Teyssier R., Carollo C. M., 2010, MNRAS, 405, 274

\bibitem[{Hao et al. (2009)}]{hao}
 Hao J., et al., 2009, ApJ, 702, 745

\bibitem[{Heller et al. (2007)}]{hell07}
 Heller C. H., Shlosman I., Athanassoula E., 2007, ApJ, 671, 226

\bibitem[{Jeeson-Daniel et al. (2011)}]{ourcompetitors}
 Jeeson-Daniel A., Dalla Vecchia C., Haas M. R., Schaye J., 2011, MNRAS, submitted

\bibitem[{Jing \& Suto (2002)}]{js02}
 Jing Y. P, Suto Y, 2002, ApJ, 574, 538

\bibitem[{Kang et al. (2007)}]{kang}
 Kang X., van den Bosch F. C., Yang X., Mao S., Mo H. J., Li C., Jing Y. P., 2007, MNRAS, 378, 1531

\bibitem[{Kasun \& Evrard (2005)}]{ke05}
 Kasun S. F., Evrard A. E., 2005, ApJ, 629, 781

\bibitem[{Kazantzidis et al. (2004)}]{kaz04}
 Kazantzidis S., Kravtsov A. V., Zentner A. R., Allgood B., Nagai D., Moore B., 2004, ApJ, 611, L73

\bibitem[{Kelly \& McKay (2004)}]{km04}
 Kelly B. C., McKay T. A., 2004, AJ, 127, 625

\bibitem[{Knebe et al. (2011)}]{MAD}
 Knebe A., et al., 2011, MNRAS, submitted (arXiv: 1104.0949)

\bibitem[{Kuhlen et al. (2007)}]{kuhlen}
 Kuhlen M., Diemand J., Madau P., 2007, ApJ, 671, 1135

\bibitem[{Macci{\`o} et al.(2003)}]{2003ApJ...588...35M} 
 Macci{\`o} A.~V., Murante G., Bonometto S.~A., 2003, ApJ, 588, 35 

\bibitem[{Maccio et al. (2007)}]{2007MNRAS.378.55M}
 Macci{\`o} A. V., Dutton A. A., van den Bosch F. C., Moore B., Potter D., Stadel J., 2007, MNRAS, 378, 55

\bibitem[{Macci{\`o} et al.(2008)}]{2008MNRAS.391.1940M}
 Macci{\`o}, A.~V., Dutton, A.~A., van den Bosch, F.~C., 2008, \mnras, 391, 1940 

\bibitem[{Madgwick et al. (2003)}]{madgwick}
 Madgwick D. S., Somerville R., Lahav O., Ellis R., 2003, MNRAS, 343, 871

\bibitem[{Mainini et al. (2003)}]{mainini}
 Mainini R., Macci{\`o} A. V., Bonometto S. A., Klypin A., 2003, ApJ, 599, 24

\bibitem[{McGurk et al. (2010)}]{mcgurk}
 McGurk R. C., Kimball A. E., Ivezi\'{c} Z., 2010, AJ, 139, 1261

\bibitem[{Moster et al. (2010)}]{moster}
 Moster B. P., Somerville R. S., Maulbetsch C., van den Bosch F. C., Macci{\`o} A. V., Naab T., Oser L., 2010, ApJ, 710, 903

\bibitem[{Munoz-Cuartas et al. (2010)}]{munoz}
 Mu\~{n}oz-Cuartas J. C., Macci\`{o} A. V., Gottl\"{o}ber, Dutton A. A., 2011, MNRAS, 411, 584 

\bibitem[{Murtagh \& Heck (1987)}]{murtagh}
 Murtagh F., Heck A., 1987, Multivariate Data Analysis (Dordrecht: Reidel)

\bibitem[{Navarro, Frenk \& White (1997)}]{nfw97}
 Navarro J. F., Frenk C. S., White S. D. M., 1997, ApJ, 490, 493

\bibitem[{Neto et al. (2007)}]{neto}
 Neto A. F., et al., 2007, MNRAS, 381, 1450


\bibitem[{Ragone-Figueroa et al. (2010)}]{rf10}
 Ragone-Figueroa C., Plionis M., Merch\'{a}n M., Gottl\"{o}ber S., Yepes G., 2010, MNRAS, 407, 581 

\bibitem[{Rogers et al. (2007)}]{rogers}
 Rogers B., Ferreras I., Lahav O., Bernardi M., Kaviraj S., Yi S. K., 2007, MNRAS, 382, 750

\bibitem[{Scarlata et al. (2007)}]{scarlata}
 Scarlata C., et al., 2007, ApJS, 172, 406

\bibitem[{Sharma \& Steinmetz (2005)}]{ss05}
 Sharma S., Steinmetz M., 2005, ApJ, 628, 21

\bibitem[{Shaw et al. (2006)}]{shaw}
 Shaw L. D., Weller J., Ostriker J. P., Bode P., 2006, ApJ, 646, 815

\bibitem[{Sheth \& Tormen (2004)}]{st04}
 Sheth R. K., Tormen G., 2004, MNRAS, 350, 1385

\bibitem[{Skibba et al. (2006)}]{sscs06} 
 Skibba R. A., Sheth R. K., Connolly A. J., Scranton R., 2006, MNRAS, 369, 68

\bibitem[{Skibba \& Sheth (2008)}]{ss08}
 Skibba R. A., Sheth R. K., 2009, MNRAS, 392, 1080 

\bibitem[{Skibba et al. 2009b}]{BHGpaper}
 Skibba R. A., van den Bosch F. C., Yang X., More S., Mo H.J., Fontanot F., 2011, MNRAS, 410, 417 

\bibitem[{Somerville et al. (2008)}]{somer}
 Somerville R. S., Hopkins P. F., Cox T. J., Robertson B. E., Hernquist L., 2008, MNRAS, 391, 481

\bibitem[{Stadel (2001)}]{pkdgrav}
 Stadel J. G., 2001, Ph.D. Thesis, University of Washington

\bibitem[{Tinker et al. (2007)}]{t07}
 Tinker J. L., Conroy C., Norberg P., Patiri S. G., Weinberg D. H., Warren M. S., 2008a, ApJ, 686, 53 

\bibitem[{Tinker et al. (2008b)}]{t08}
 Tinker J., Kravtsov A. V., Klypin A., Abazajian K., Warren M., Yepes, G., Gottl\"{o}ber S., Holz D. E., 2008b, ApJ, 688, 709

\bibitem[{van den Bosch (2002)}]{vdB02b}
 van den Bosch F. C., 2002, MNRAS, 332, 456

\bibitem[{van den Bosch et al. (2002)}]{vdB02}
 van den Bosch F. C., Abel T., Croft R. A. C., Hernquist L., White S. D. M., 2002, ApJ, 576, 21

\bibitem[{van den Bosch et al. (2005)}]{vdB05c}
 van den Bosch F. C., Weinmann S. M., Yang X., Mo H. J., Li C., Jing Y. P., 2005, MNRAS, 361, 1203 

\bibitem[{van den Bosch et al. (2007)}]{vdB07}
 van den Bosch F. C., Yang X., Mo H. J., Weinmann S. M., Macci\`{o} A. V., More S., Cacciato M., Skibba R., Kang X., 2007, MNRAS, 376, 841

\bibitem[{Wang et al. (2007)}]{wang07}
 Wang H. Y., Mo H. J., Jing Y. P., 2007, MNRAS, 375, 633

\bibitem[{Wang et al. (2008)}]{wang08}
 Wang Y., Yang X., Mo H. J., Li C., van den Bosch, F. C., Fan Z., Chen X., 2008, MNRAS, 385, 1511

\bibitem[{Wang et al. (2010)}]{wangh10}
 Wang H., Mo H. J., Jing Y. P., Yang X., Wang Y., 2011, MNRAS, 413, 1973


\bibitem[{Wechsler et al. (2006)}]{w05}
 Wechsler R. H., Zentner A. R., Bullock J. S., Kravtsov A. V., Allgood B., 2006, ApJ, 652, 71

\bibitem[{Wetzel et al. (2007)}]{ww07}
 Wetzel A. A., Cohn J. D., White M., Holz D. E., Warren M. S., 2007, ApJ, 656, 139

\bibitem[White \& Rees(1978)]{1978MNRAS.183..341W}
 White, S.~D.~M., Rees, M.~J., 1978, \mnras, 183, 341 

\bibitem[{Woo et al. (2008)}]{woo}
 Woo J., Courteau S., Dekel A., 2008, MNRAS, 390, 1453

\bibitem[{Yang et al. (2006)}]{yang}
 Yang X., van den Bosch F. C., Mo H. J., Mao S., Kang X., Weinmann S. M., Guo Y., Jing Y. P., 2006, MNRAS, 369, 1293

\bibitem[{Yip et al. (2004)}]{yip}
 Yip C. W., et al., 2004, AJ, 128, 2603

\bibitem[{Zehavi et al. (2005)}]{z05}
 Zehavi I., et al., 2005, ApJ, 630, 1

\bibitem[{Zemp et al. (2009)}]{zemp}
 Zemp M., Diemand J., Kuhlen M., Madau P., Moore B., Potter D., Stadel J., Widrow L., 2009, MNRAS, 394, 641

\bibitem[{Zentner et al. (2005a)}]{zentner05a}
 Zentner A. R., Berlind A. A., Bullock J. R., Kravtsov A. V., Wechsler R. H., 2005a, ApJ, 624, 505

\bibitem[{Zentner et al. (2005b)}]{zentner05b}
 Zentner A. R., Kravtsov A. V., Gnedin O. Y., Klypin A. A., 2005b, ApJ, 629, 219

\end{thebibliography}
\end{document}